\documentclass[article]{elsarticle}
\makeatletter
\def\ps@pprintTitle{%
	\let\@oddhead\@empty
	\let\@evenhead\@empty
	\def\@oddfoot{\centerline{\thepage}}%
	\let\@evenfoot\@oddfoot}
\makeatother

\usepackage{lineno,hyperref} 
\usepackage{threeparttable}
\usepackage{float}
\usepackage{booktabs}
\usepackage{amsmath}
\usepackage{comment}
\usepackage{color}

\usepackage{amsfonts}
\usepackage{amssymb}
\usepackage{amsmath}
\usepackage{amsthm}

\usepackage{natbib}

\modulolinenumbers[5]
\newcommand{\comm}[1]{}
\begin{document}

\begin{frontmatter}

\title{Atomistic and mean-field estimates of effective stiffness tensor\\ of nanocrystalline copper}

\author[]{Katarzyna Kowalczyk-Gajewska\corref{mycorrespondingauthor}}
\ead{kkowalcz@ippt.pan.pl}

\author{Marcin Ma\'zdziarz }
\ead{mmazdz@ippt.pan.pl}

\address{Institute of Fundamental Technological Research Polish Academy of Sciences,
Pawi\'nskiego 5B, 02-106 Warsaw, Poland}
\cortext[mycorrespondingauthor]{Corresponding author}





\begin{abstract}
The full elasticity tensor for nano-crystalline copper is derived in molecular simulations by performing numerical tests for a set of {generated samples} of the polycrystalline material. The results are analysed with respect to the anisotropy degree of the overall stiffness tensor resulting from the limited number of grain orientations and their spatial distribution. The dependence of the overall bulk and shear moduli of an isotropized polycrystal on the average grain diameter is analysed. It is found that while the shear modulus decreases with  grain size, the bulk modulus shows negligible dependence on the grain diameter and is close to the bulk modulus of a single crystal. A closed-form mean-field model of effective elastic properties for a bulk nano-grained polycrystal with cubic grains, i.e. made of a material with cubic symmetry, is formulated. In the model all parameters are based on the data for a single crystal and on the averaged grain size without any need for additional fitting. It is shown that the proposed model provides predictions of satisfactory qualitative and quantitative agreement with atomistic simulations.
\end{abstract}


\end{frontmatter}


\section{Introduction}
\label{sec:Int}

Metal polycrystals are examples of heterogeneous materials in which heterogeneity results mainly from different orientations of local anisotropy axes in each grain. When the size of grains is sufficiently large the continuum mechanics framework is applicable and usually the boundaries between the grains are treated as perfect interfaces that do not require special treatment.

In order to estimate effective properties of a coarse-grained polycrystalline material a micromechanical methodology is used. Besides the classical Voigt and Reuss estimates, the self-consistent model due to Kroner \cite{Kroner58} and Hill \cite{Hill65} is most often employed. This methodology was successfully extended to the (visco)plastic crystals and large strain regime \cite{Molinari87,Lebensohn93}. Verification of such micromechanical mean-field estimates is usually performed by means of full-field simulations employing either the finite element method (FEM) \cite{Kamaya20092642,Fan20105008} or the fast Fourier transform (FFT) technique \cite{Prakash09,Liu10,Lebensohn12}.

Nanocrystalline metals are polycrystalline materials with a grain size smaller than 100 nm \cite{Gleiter00,Gao13}. In such materials two phases can be distinguished: grain cores and grain boundaries. The effect of the grain boundaries on the overall properties of a bulk material increases with decreasing grain size \cite{Sanders97,Gao13}. At the macroscopic level a nanograined material is still described in the framework of continuum theories. Since there is a need for quick and simple estimates of effective properties, which use only basic knowledge of the material behaviour: single grain properties, orientation distribution (texture) within the material volume and the averaged grain size, micromechanical mean-field estimates are also utilized in this case. Most of the mean-field schemes proposed for nanocrystalline materials describe them as composite media made of two or more phases. One phase is the grain core and the other phase (or phases) represents the behaviour of the grain boundary. In the early mixture-based model \cite{Carsley95} a nanophase metal was described as a mixture of a bulk intergranular region and a grain boundary built of amorphous metal. The overall strength of the material was obtained as a simple volume average of the strengths of the two phases. In \cite{Kim99} a refined mixture model was developed in which the intercrystalline phase was composed of three sub-phases: grain boundary, triple line junctions and quadratic nodes. Additionally pores were considered. All phases were modelled as isotropic and in order to obtain overall elastic moduli the implicit self-consistent relation was used.  The concept of grain boundary subdivision into sub-phases  was followed in \cite{Benson01,Qing06}, although using simple volume averages to obtain the overall Young's  modulus and strength, while in the approach presented by \cite{Zhou07} the overall properties were assessed by using subsequently iso-strain and iso-stress assumptions for finding the average response of the grain core and the two grain boundary sub-phases.  In \cite{Sharma03} a nano-grained material was described as a crystalline matrix with embedded ellipsoidal flat disks representing the grain boundary. To estimate the overall strength that accounts for grain boundary sliding the disks' stiffness moduli were anisotropic with one modulus vanishing, and the Mori-Tanaka (MT) scale-transition scheme was applied. 
In \cite{Jiang04} the idea of the generalized self-consistent (GSC) model was used to find the strength and stiffness of a nanograined polycrystal: the coating of the spherical grain was assumed to be made of an isotropic material representing the grain boundaries. The grain cores were assumed to be elastically isotropic, while retaining their plastic anisotropy. This idea was followed by \cite{Capolungo07} and \cite{Ramtani09}. In the former model the implicit formulae of the GSC scheme \cite{Christensen79} were replaced by the explicit relations of the Mori-Tanaka model under the assumption that the grains are embedded in the matrix of the grain boundary phase. {Similarly, in \cite{Mercier07} a two-phase elastic-viscoplastic model combined with the Taylor-Lin homogenization theory was used to find the overall yield stress of nanocrystalline copper. The core-shell modelling framework was also used in the case of nanowires \cite{Chen06}.} The main difficulty of all these models is related to a valid description of the grain boundary phase(s) and quantification of its volume fraction.

Differently than in the case of coarse-grained materials, in the case of nano-grained polycrystals the use of full-field simulations based on continuum theory for the purpose of verification or calibration of the above-discussed models is questionable. Therefore atomistic simulations based on molecular dynamics are performed to assess the elastic moduli and strength of such materials \cite{Schiotz99,Chang03,Choi12,Gao13,Mortazavi2014,Fang16}. An alternative approach is to use refined constitutive models, for example those incorporating scale effects through gradient enhancements, within the framework of FEM calculations \cite{Kim20123942}. 

In atomistic simulations of elastic properties, usually only the tensile Young's modulus is determined, see e.g. \cite{Schiotz99,Gao13}. It is observed that this modulus decreases with a decreasing grain size. This tendency is qualitatively and quantitatively reproduced by two- or multi-phase micromechanical models after proper adjustment of the grain boundary stiffness and the volume fraction, e.g. \cite{Jiang04,Ramtani09,Gao13}. 

{The goal of the present paper is to estimate the effective elastic properties of nanocrystalline copper taking  the anisotropy of a single crystallite fully into account. To this end, a series of static molecular simulations are performed to find all 21 components of the stiffness tensor and an anisotropic mean-field core-shell model of a nanograined polycrystal is formulated within the continuum mechanics framework. The results of both approaches are compared.}

The paper is constructed as follows. The next section describes the details of the performed atomistic simulations. Section \ref{sec:Cont}  
presents the proposed formulation of the two-phase model of a nano-grain polycrystal. Additionally, the isotropisation procedure is outlined for the anisotropic stiffness acquired in the simulations. Section~\ref{sec:Res} discusses the results of the atomistic simulations and compares their outcomes with the respective mean-field estimates. The paper is closed with conclusions. 

\section{Computational methods}
\label{sec:Cm}
All molecular simulations were performed using the Large-scale Atomic/Molecular Massively Parallel Simulator (LAMMPS) \cite{Plimpton1995} and  a scientific visualization and analysis software for atomistic simulation data OVITO was used to visualize and analyse simulation results \cite{Ovito2010}. Due to interest in the static behaviour of the material, the molecular statics (MS) approach at the temperature of 0K \cite{Tadmor2011,Maz2011,Maz2010} was used, which is more appropriate in such cases than molecular dynamics (MD).
Copper is metal arranged in a face-centered cubic (FCC) structure. The potentials based on the Embedded Atom Model (EAM) have been successful in the description of many metallic systems \cite{Tadmor2011}. 
For analysed FCC copper, the well known EAM potential parametrized by \cite{Mishin2001} and taken from NIST Interatomic Potentials Repository \cite{Becker2013277} was utilized. This potential reproduces the copper monocrystal FCC lattice constant  $a_{FCC}$=3.615\,\AA, the cohesive energy $E_{c}$=-3.54 eV, and the elastic constants in crystallographic axes coinciding with Cartesian coordinate system axes, $C_{1111}$=169.88\,GPa, $C_{1122}$=122.60\,GPa and  $C_{2323}$=76.19\,GPa.

Polycrystal structures were generated using the Voronoi tessellation method by the Atomsk program which is designed for creating, converting and manipulating atomic systems \cite{Hirel2015212}. Data for the positions of seeds, orientations and size distribution of grains are attached in the \textit{Supplementary material} (polycrystals.txt and polycrystals.dat files). 

Comprehensive study of the topological properties of the Voronoi tessellations of cubic systems can be found in \cite{Lucarini2009}. In physics of condensed matter, the Voronoi cell of the lattice point of a crystal is well known as the Wigner-Seitz cell, while the Voronoi cell of the reciprocal lattice point is called the Brillouin zone.
All the Voronoi tessellation cells are convex polyhedra. For the simple cubic (SC) crystal system these polyhedra are cubes (having 12 edges, 6 faces and 8 trivalent vertices) with a side length $V_{cell}^{-1/3}$ and total surface area, $A = 6\,V_{cell}^{-2/3}$, for the body-centered cubic (BCC) crystal system are truncated octahedra (having 36
edges, 14 faces and 24 trivalent vertices) of side length $\approx 0.445\,V_{cell}^{-1/3}$ and total, $A \approx 5.315\, V_{cell}^{-2/3}$ and for the face-centered cubic (FCC) crystal system these polygons are rhombic dodecahedra (having 24 edges, 12 faces, 6 trivalent vertices and 8 tetravalent vertices) of side length $\approx 0.687\,V_{cell}^{-1/3}$ and total, $A \approx 5.345\, V_{cell}^{-2/3}$. The standard isoperimetric quotient $IQ= 36\pi V_{cell}^{2}/A^3$,
non-dimensional measure of the surface-to-volume ratio of a solid ($IQ$ = 1 for a sphere), is 0.524, 0.753, and 0.74 for the SC, BCC and FCC system, respectively. Thus, $IQ$ decreases in value with increasing shape irregularity.

To examine the effect of the number and size of grains on mechanical properties of a polycrystalline material, eleven distinct copper samples were considered in this work: monocrystal, polycrystals of different sizes from 25$^3$ to 100$^3$ copper monocrystal unit cells with fixed number of 2x4$^3$ grains in BCC arrangement, a polycrystals with fixed size of 50$^3$ copper monocrystal unit cells with 2x2$^3$, 2x3$^3$, 2x4$^3$ and 2x5$^3$ grains in BCC arrangement, a polycrystal with fixed size of 50$^3$ copper monocrystal unit cells with 4x3$^3$ grains in FCC arrangement, a polycrystal of the same size and 5$^3$ grains in random arrangement and additionally polycrystal with a fixed size of 100$^3$ copper monocrystal unit cells with 2x2$^3$ grains in BCC arrangement, see Tab.\ref{tab:Samples} and Fig.\ref{fig:Samples}.

Structures created in such a geometrical way usually are not as dense as those observed in real materials. 
In order to make the initial configurations more physical and to move atoms at grain boundaries to their energetically favourable positions the generated structures  were prerelaxed. 
Energy minimisation with a periodic boundary condition applied to all facets of the sample
was carried out with the non-linear conjugate gradient algorithm. Convergence was achieved when the
relative change in energy and forces, between two consecutive iterations  were less than $10^{-8}$. After energy minimization process an external pressure equal to \textit{zero} was applied to the simulation box. This procedure allows the simulation box shape and volume to vary during the iterations of the minimizer so that the final configuration should be both an energy minimum for the potential energy of the atoms, as well as the system pressure tensor should be close to the prescribed external tensor \cite{Plimpton1995}.
The elastic constants, of all relaxed structures, $\bar{C}_{ijkl}$ were computed with the strain-stress approach with the maximum strain amplitude set to be 10$^{-4}$ \cite{Plimpton1995,Mazdziarz2015}.

There exist many computational analysis approaches to classify local atomic arrangements in computer-based large-scale atomistic simulations of crystalline solids that can be used to characterize when the atom is part of a perfect lattice, a local defect (e.g. a stacking fault or dislocation), or is positioned on a surface. The most common techniques include popular neighbor analysis (CNA), centrosymmetry parameter analysis (CPA), coordination analysis (CA), energy filtering, etc., see \cite{Stukowski2012}. We have found that in our case the cohesive energy/atom quantity and {CNA} are very useful techniques to analyse polycrystals.

\section{Continuum mechanics estimates of elastic moduli}
\label{sec:Cont}

Overall elastic properties of a coarse-grained polycrystalline material are usually found employing the micromechanics methodology set in a continuum mechanics framework. 
According to the micromechanical theories polycrystals are one-phase heterogeneous materials in which heterogeneity roots from the different orientation  of crystallographic axes 
from grain-to-grain in the representative volume of polycrystalline continuum. Effective properties are assessed assuming that local elastic properties are known and then performing a micro-macro transition. The standard estimates for one-phase polycrystalline materials 
{are collected in \ref{Ap:A}}. 
None of these estimates are sensitive to the grain size.

\begin{figure}[h!]
	\centering
	\includegraphics[width=.7\textwidth]{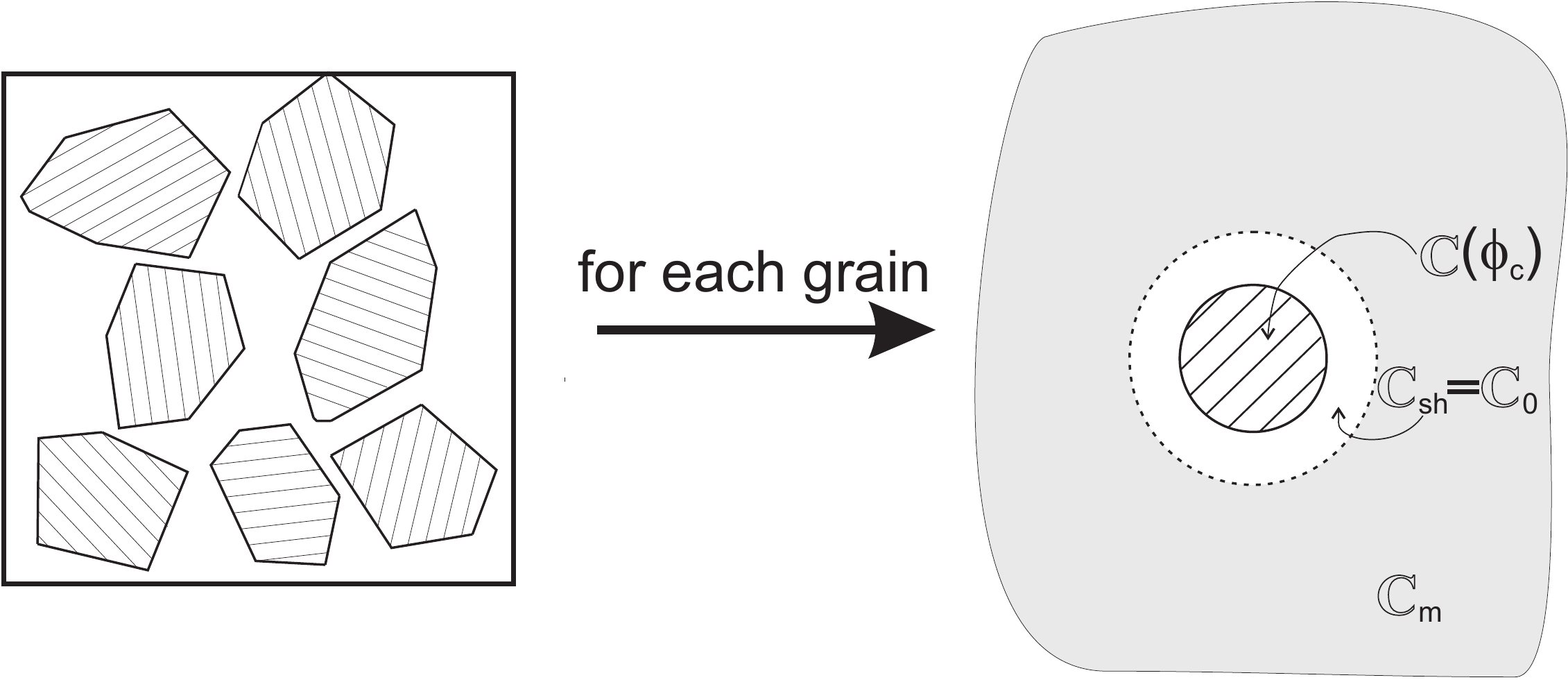}
	\caption{Schematic of the core-shell model of the nanograin polycrystal.}
	\label{fig:CoreShell}
\end{figure}

As discussed in Sec.\ref{sec:Int}, in order to account for the size effect, which is present in nano-crystalline aggregates, a family of models using the concept of a two- or multi-phase polycrystal were proposed. Two such models are described below that are called \emph{the MT and SC core-shell models}, respectively. The presented formulation follows the idea that has been proposed by \cite{Jiang04} and \cite{Capolungo07}. According to this approach an additional phase that represents a transient zone between the grains is introduced. This phase forms a coating of a specified thickness. An impact of this transient zone on the effective properties is more significant for smaller grains.

In opposition to the mentioned studies \cite{Jiang04,Capolungo07}, in order to estimate the overall stiffness tensor the elastic anisotropy of the grain core is thoroughly taken into account. The local constitutive relation between the stress tensor
$\boldsymbol{\sigma}$ and strain tensor
$\boldsymbol{\varepsilon}$ in the grain core is
\begin{equation}\label{locconst}
\boldsymbol{\sigma}=\mathbb{C}(\phi_c)\cdot\boldsymbol{\varepsilon},\quad
\boldsymbol{\varepsilon}=\mathbb{S}(\phi_c)\cdot\boldsymbol{\sigma},\quad\mathbb{S}=\mathbb{C}^{-1}\,,
\end{equation}
where $\mathbb{C}(\phi_c)$ and $\mathbb{S}(\phi_c)$ are the fourth order anisotropic elastic stiffness and compliance
tensors, respectively, and $\phi^c$ denotes orientation of local axes
$\{\mathbf{a}_k\}$ with respect to some macroscopic frame
$\{\mathbf{e}_k\}$ specified by e.g. three Euler angles. On the other hand the grain shell is isotropic with the stiffness tensor equal to the lower zeroth-order bound $\mathbb{C}_0$ (see  \ref{Ap:A}). The latter specification is based on the results of atomistic simulations reported in Sec.\ref{ssec:ResAS} \comm{the next section}.  Macroscopic relations for the averaged
fields $\mathbf{E}=\langle \boldsymbol{\varepsilon}\rangle $ and
$\boldsymbol{\Sigma}=\langle \boldsymbol{\sigma}\rangle $ in the polycrystal are then as follows
\begin{equation}
\boldsymbol{\Sigma}=\bar{\mathbb{C}}\cdot\mathbf{E},\quad
\mathbf{E}=\bar{\mathbb{S}}\cdot\boldsymbol{\Sigma},\quad
\bar{\mathbb{S}}=\bar{\mathbb{C}}^{-1}\,,
\end{equation}
where averaging is performed over the representative material volume so that $\langle . \rangle =\frac{1}{V}\int_V(.) dV $ and $\bar{\mathbb{C}}$ ($\bar{\mathbb{S}}$) is the effective stiffness (compliance) tensor of the polycrystal to be found by the mean-field model.

The effective stiffness $\bar{\mathbb{C}}$ is estimated by assuming that the coated grain is an inhomogeneity embedded in the infinite matrix of the stiffness $\mathbb{C}_m$ and using the methodology of the double-inclusion model \cite{HoriNematNasser93}. As a result it is obtained
\begin{equation}\label{eq:core-shell}
\bar{\mathbb{C}}_{\rm{CS}}=\left[f_0\mathbb{C}_0\mathbb{A}+(1-f_0)\left<\mathbb{C}(\phi^c)\mathbb{A}(\phi^c)\right>_{\mathcal{O}}\right]\left[f_0\mathbb{A}_0+(1-f_0)\left<\mathbb{A}(\phi^c)\right>_{\mathcal{O}}\right]^{-1}
\end{equation}						
where
\begin{equation}
\mathbb{A}(\phi^c)=(\mathbb{C}(\phi^c)+\mathbb{C}_*(\mathbb{C}_{\rm{m}}))^{-1}(\mathbb{C}_{\rm{m}}+\mathbb{C}_*(\mathbb{C}_{\rm{m}}))\,,
\end{equation}	
\begin{equation}
\mathbb{A}_0=(\mathbb{C}_0+\mathbb{C}_*(\mathbb{C}_{\rm{m}}))^{-1}(\mathbb{C}_{\rm{m}}+\mathbb{C}_*(\mathbb{C}_{\rm{m}}))\,,
\end{equation}
and $f_0$ is the volume fraction of the transient zone,  $\mathbb{C}_*(\mathbb{C}_{\rm{m}})$ is the Hill tensor \cite{Hill65} depending on the stiffness $\mathbb{C}_{\rm{m}}$ of matrix
material and the shape of the coated grain (here assumed to be spherical) and  $\langle . \rangle_{\mathcal{O}}$ denotes the averaging over grain crystallographic orientations. Two variants of the core-shell model are considered. For the first variant $\mathbb{C}_{\rm{m}}=\mathbb{C}_0$ is assumed (the infinite matrix has shell properties), while for the second variant $\mathbb{C}_{\rm{m}}=\bar{\mathbb{C}}_{\rm{CS}}$ (the infinite matrix has the effective properties to be found and formula (\ref{eq:core-shell}) is implicit). Due their correspondence to the Mori-Tanaka (MT) method and the self-consistent (SC) model, widely used for the two-phase composites, these two variants are referred as MT and SC core-shell models, respectively. Note that for a coarse-grained polycrystal ($f_0\rightarrow 0$) the effective properties $\bar{\mathbb{C}}_{\rm{CS/MT}}$ and $\bar{\mathbb{C}}_{\rm{CS/SC}}$ approach the Hashin-Shtrikman lower bound $\bar{\mathbb{C}}_{\rm{HS}}$ (Eq. \ref{HSbound}) and the self-consistent estimate $\bar{\mathbb{C}}_{\rm{SC}}$ (Eq. \ref{sc1}) for a one-phase polycrystal, respectively. The latter result is usually recommended in the literature \cite{Walpole81}. For very small grains ($f_0\rightarrow 1$) the estimates approach each other and coincide with $\mathbb{C}_0$. Consequently, for the same $f_0$ the stiffness moduli predicted by the MT variant of the core-shell model will be always lower then those predicted by the SC variant and the gap between estimates will increase with a grain size.

Two cases are analysed. The first one for which the finite set of $N_g$ grain orientations is known, and the second one for which an infinite set of orientations of random uniform distribution in the orientation space \cite{Bunge} is assumed. The latter case results in exactly isotropic overall properties of polycrystalline material. It should be highlighted that in both cases it is assumed that the volume fraction of grains with a specified orientation is the same and that the grains are equiaxed so that the shape of a grain core and its coating can be approximated as spherical.

For two analysed cases of orientation distribution, i.e. discrete and infinite, the averaging operation $\langle . \rangle_{\mathcal{O}}$ present in Eq. (\ref{eq:core-shell}) reduces to \cite{Bunge}:
\begin{equation}\label{Eq:orient}
	\langle . \rangle_{\mathcal{O}}=\frac{1}{N_g}\sum_{i=1}^{N_g}(.)_i\quad\textrm{or}\quad
	\langle . \rangle_{\mathcal{O}}=\frac{1}{8\pi^2}\int_0^{2\pi}\int_0^{\pi}\int_0^{2\pi}(\quad\cdot\quad)\sin\psi
	\mathrm{d}\varphi_1\mathrm{d}\psi\mathrm{d}\varphi_2\,,
\end{equation}
respectively. In the second case, for the averaging over the fourth order tensor field $\mathbb{T}(\phi^c)$, it can be simplified to \cite{Forte96,Rychlewski01,Kowalczyk09}:
\begin{displaymath}
	\langle \mathbb{T}(\phi_c) \rangle_{\mathcal{O}} = \frac{1}{3}T(\phi_c)_{iijj}\mathbb{I}^{\rm{P}}+\frac{1}{5}\left(T(\phi_c)_{ijij}-\frac{1}{3}T(\phi_c)_{iijj}\right)\mathbb{I}^{\rm{D}}\,,
\end{displaymath}
where $\mathbb{I}^{\rm{P}}$ and $\mathbb{I}^{\rm{D}}$ are fourth order orthogonal projectors to the hydrostatic and deviatoric subspaces of the second order tensor space. In any orthonormal basis they have components: $I^{\rm{P}}_{ijkl}=1/3\delta_{ij}\delta_{kl}$ and $I^{\rm{D}}_{ijkl}=1/2(\delta_{ik}\delta_{jl}+\delta_{il}\delta_{jk})-I^{\rm{P}}_{ijkl}$. The scalars $T(\phi_c)_{iijj}$ and $T(\phi_c)_{ijij}$ are invariants of the fourth order tensor, such that their values are independent of the selected basis, therefore they can be calculated for the tensor representations in the local crystal frame $\{\mathbf{a}_k\}$. The overall stiffness tensor is then isotropic and specified by two parameters: the bulk modulus $\bar{K}$ and shear modulus $\bar{G}$; namely
\begin{displaymath}
	\bar{\mathbb{C}}=3\bar{K}\mathbb{I}^{\rm{P}}+2\bar{G}\mathbb{I}^{\rm{D}}\,.
\end{displaymath}  

Eq. (\ref{eq:core-shell}) simplifies further if the crystal has locally cubic lattice symmetry. In such a case, the local stiffness tensor can be expressed in the following spectral form \cite{Walpole81,Rychlewski95,Kowalczyk09}
\begin{equation}\label{Eq:SpekCubic}
	\mathbb{C}(\phi^c)=3K\mathbb{I}^{\rm{P}}+2G_1(\mathbb{K}(\phi^c)-\mathbb{I}^{\rm{P}})+2
	G_2(\mathbb{I}-\mathbb{K}(\phi^c))\,,
\end{equation}
where
\begin{equation}
	\mathbb{K}=\sum_{k=1}^3\mathbf{a}_k\otimes\mathbf{a}_k\otimes\mathbf{a}_k\otimes\mathbf{a}_k\,,
\end{equation}
and $3K$, $2G_1$ and $2G_2$ are Kelvin moduli specified as
\begin{equation}\label{Eq:moduli}
	3K=C_{1111}+2C_{1122}\,,\quad 2G_1=C_{1111}-C_{1122}\,,\quad 2G_2=2C_{2323}\,,
\end{equation}
and $C_{ijkl}$ are the components of the local stiffness tensor in the basis $\mathbf{a}_k$. 

For the second case of orientation distribution in the representative volume, it can be easily verified that all discussed estimates for the polycrystal bulk modulus are equal: $\bar{K}=K$.

With regard to the shear modulus the respective expressions for classical estimates are collected in Table \ref{tab:LinearEstimates}. The zeroth-order lower and upper bounds for $\bar{G}$ are equal to $\min{(G_1,G_2)}$ and $\max{(G_1,G_2)}$, respectively, such that the $\bar{G}$-estimate by the MT core-shell model for random orientation distribution, is specified as
\begin{equation} 
{{\bar{G}_{\rm{CS/MT}}=G_1\frac{4G_1(4 G_1 + 21G_2) + 3K(6 G_1 + 19 G_2) -
			3(G_2 - G_1) (8 G_1 + 9 K) f_0}{
			4G_1(19 G_1 + 6 G_2) + 3K(21 G_1 + 4 G_2) + 
			18 (G_2 - G_1) (2 G_1 + K) f_0}}}
\end{equation}
for $G_1<G_2$, or,
\begin{equation}
{{\bar{G}_{\rm{CS/MT}}=G_2 \frac{4G_2(19 G_1 + 6G_2) + 3K(16 G_1 + 9 G_2) - 
			2(G_1 - G_2) (8 G_2 + 9 K) f_0}{
			4G_2(9 G_1 + 16 G_2) + 3K(6 G_1 + 19 G_2) + 
			12 (G_1 - G_2) (2 G_2 + K) f_0}}}
\end{equation}		
for $G_1>G_2$. The $\bar{G}$-estimate by the SC core-shell model, again for random orientation distribution, is obtained as a solution of the following third order equation
\begin{equation}\label{G-CS-SC-1}
\begin{split}
8\bar{G}^3_{\rm{CS/SC}}+[9K+4G_1+12f_0(G_2-G_1)]\bar{G}^2_{\rm{CS/SC}}+\qquad\qquad\quad\quad\\
-[3G_2(4G_1+K) - 9 f_0 K(G_2-G_1)]\bar{G}_{\rm{CS/SC}}-6 G_1 G_2 K=0
\end{split}
\end{equation}
for $G_1<G_2$ or
\begin{equation}\label{G-CS-SC-2}
\begin{split}
8\bar{G}^3_{\rm{CS/SC}}+[9K+4G_1+8f_0(G_1-G_2)]\bar{G}^2_{\rm{CS/SC}}+\qquad\qquad\quad\quad\\
-[3G_2(4G_1+K) - 6 f_0 K(G_1-G_2)]\bar{G}_{\rm{CS/SC}}-6 G_1 G_2 K=0
\end{split}
\end{equation}
for $G_1>G_2$. By analysing the coefficients of the cubic polynomials (\ref{G-CS-SC-1}) and (\ref{G-CS-SC-2}) it can be easily proved that they have a single positive root, such that the solution is unique. It can be specified by a closed formula using  known relations for roots of the cubic polynomial.

In the case of polycrystalline materials composed of a finite set of $N_g$ grains with randomly selected orientations, the overall stiffness tensor can be estimated by either of Eqs. (\ref{lowerL}-\ref{sc1}) or Eq. (\ref{eq:core-shell}) and is in general anisotropic. With an increasing number of orientations the overall properties approach isotropy. The discrepancy between the current stiffness obtained for a given set of orientations and its isotropic limit can be quantified using some anisotropy measure. Two definitions of the anisotropy factor used in this study result from the spectral decomposition of the overall stiffness.
The first anisotropy measure $\zeta_1$ is motivated by the Zener parameter proposed for cubic crystals \cite{zener1948elasticity}. This is defined as
\begin{equation}\label{Eq:zeta1}
	\zeta_1=\frac{\rm{min}(\lambda_K)}{\rm{max}(\lambda_K)}\leq 1\,,\quad K=2,\ldots,6\,,
\end{equation}
where $\lambda_K$ are the eigenvalues (the Kelvin moduli) of $\bar{\mathbb{C}}$ obtained by means of its spectral decomposition \cite{Cowin95,Rychlewski95} and ordered according to the value of the invariant $P_{Kiijj}$ of the corresponding eigenprojector $\mathbb{P}_K$ \cite{Kowalczyk09}. Note that $\zeta_1$ is equal to 1 for isotropic $\bar{\mathbb{C}}$ and equal to the Zener parameter for cubic $\bar{\mathbb{C}}$ if for such $\bar{\mathbb{C}}$ we have $G_1>G_2$ (see Eq. \ref{Eq:SpekCubic}). The fulfilment of $\zeta_1=1$ is necessary to ensure elastic isotropy, however, it is not sufficient in the general-anisotropy case (it is sufficient for volumetrically isotropic materials \cite{Kowalczyk12}). 

The second anisotropy measure $\zeta_2$ is defined by a norm of difference between the closest isotropic approximation of $\bar{\mathbb{C}}$ and the actual $\bar{\mathbb{C}}$. The difference is normalized by the norm of $\bar{\mathbb{C}}$. The closest isotropic approximation of $\bar{\mathbb{C}}$ is established employing the Log-Euclidean metric as proposed in \cite{Moakher06}. The anisotropy factor is then calculated as \cite{Kowalczyk11b}
\begin{equation}\label{Eq:zeta2}
	\zeta_2=\frac{||\mathrm{Log}\bar{\mathbb{C}}-\mathrm{Log}\bar{\mathbb{C}}^{\mathcal{L}}_{\rm{iso}}||}{||\mathrm{Log}\bar{\mathbb{C}}||}\geq 0\,,
\end{equation}
where $||\mathbb{A}||=\sqrt{A_{ijkl}A_{ijkl}}$ and $\mathrm{Log}\mathbb{A}=\sum_{K}\mathrm{log}\lambda_L\mathbb{P}_K$ ($\lambda_K$ - eigenvalues of $\mathbb{A}$, $\mathbb{P}_K$ - eigenprojectors of $\mathbb{A}$, both resulting from its spectral decomposition \cite{Kowalczyk09}), and
\begin{equation}
	\bar{\mathbb{C}}^{\mathcal{L}}_{\rm{iso}}=\exp\left(\frac{1}{3}Z_{iijj}\right)\mathbb{I}^{\rm{P}}+\exp\left(\frac{1}{5}\left(Z_{ijij}-\frac{1}{3}Z_{iijj}\right)\right)\mathbb{I}^{\rm{D}}\,\quad\textrm{for}\quad
	\mathbb{Z}=\mathrm{Log}\bar{\mathbb{C}}\,.
\end{equation}
Naturally, $\zeta_2$ is equal to 0 for isotropic $\bar{\mathbb{C}}$. The fulfilment of $\zeta_2=0$ is sufficient to ensure elastic isotropy. In further analysis of results the following scalars:
\begin{equation}\label{Eq:IsoBulkShear}
	\bar{K}^{\mathcal{L}}_{\rm{iso}}=\frac{1}{3}\exp\left(\frac{1}{3}Z_{iijj}\right)\,,\quad
	\bar{G}^{\mathcal{L}}_{\rm{iso}}=\frac{1}{2}\exp\left(\frac{1}{5}\left(Z_{ijij}-\frac{1}{3}Z_{iijj}\right)\right)
\end{equation}
will be used as estimates of isotropic bulk and shear moduli corresponding to, in general, an anisotropic estimate of effective stiffness tensor obtained for a finite set of $N_g$ grain orientations. 

It should be stressed that for a polycrystal of cubic grains (i.e. made of cubic symmetry material) all classical continuum-mechanics (\ref{lowerL}-\ref{sc1}) and core-shell model estimates (\ref{eq:core-shell}) -- if shell properties are defined as zeroth-order lower bound -- deliver exactly the same value of the isotropic bulk modulus: $\bar{K}^{\mathcal{L}}_{\rm{iso}}=K$, independently of the number of orientations \cite{Walpole85}.

\section{Results}
\label{sec:Res}

\subsection{Results of atomistic simulations}
\label{ssec:ResAS}

The computational samples analysed by atomistic approach are denoted as $$N_{\rm{UC}}-N_g-\rm{SYS}$$ where $N_{\rm{UC}}$ is a number of unit cells, $N_g$ - a number of grain orientations and $\rm{SYS}$ denotes the system of grain distribution, i.e.: BCC, FCC or random, see Table \ref{tab:Samples}.
The finite set of $N_g$ orientations has been considered, namely polycrystalline representative volumes composed of copper grains with 16, 54, 128 or 250 randomly selected orientations have been analysed. The orientation was defined in terms of three Euler angles. The random selection of a set of Euler angles was done in a standard way, i.e. assuming uniform distribution of orientations in the orientation  space and taking into account the specific features of the non-Euclidean space of Euler angles \cite{Bunge}. 
Elasticity tensors $\bar{\mathbb{C}}$\, derived from molecular simulations of analysed samples are listed in the Table \ref{tab:Cij}.

\begin{figure}[H]
	\centering
	\begin{tabular}{ccc}
			\includegraphics[width=0.34\linewidth]{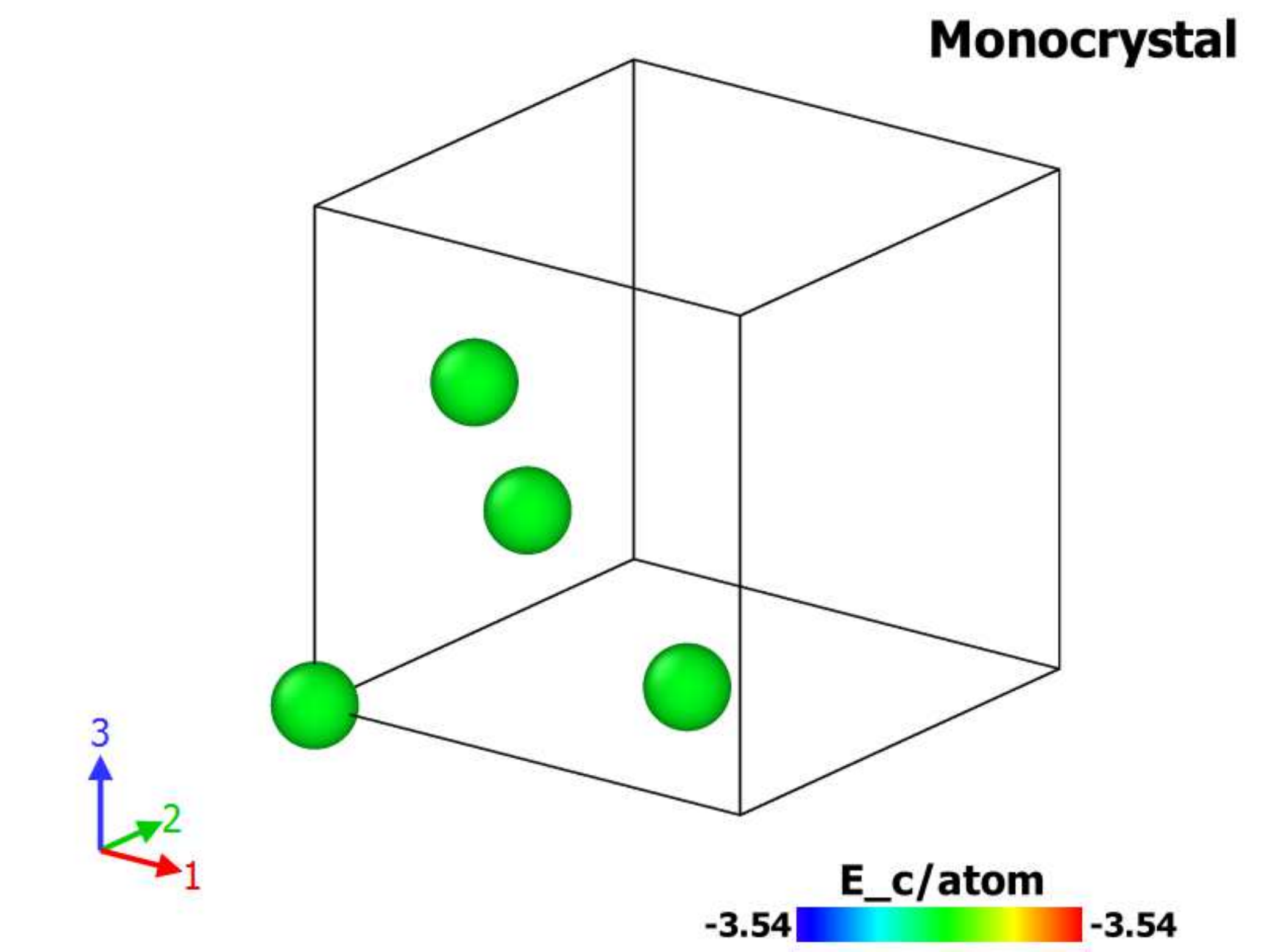} &
		\includegraphics[width=0.34\linewidth]{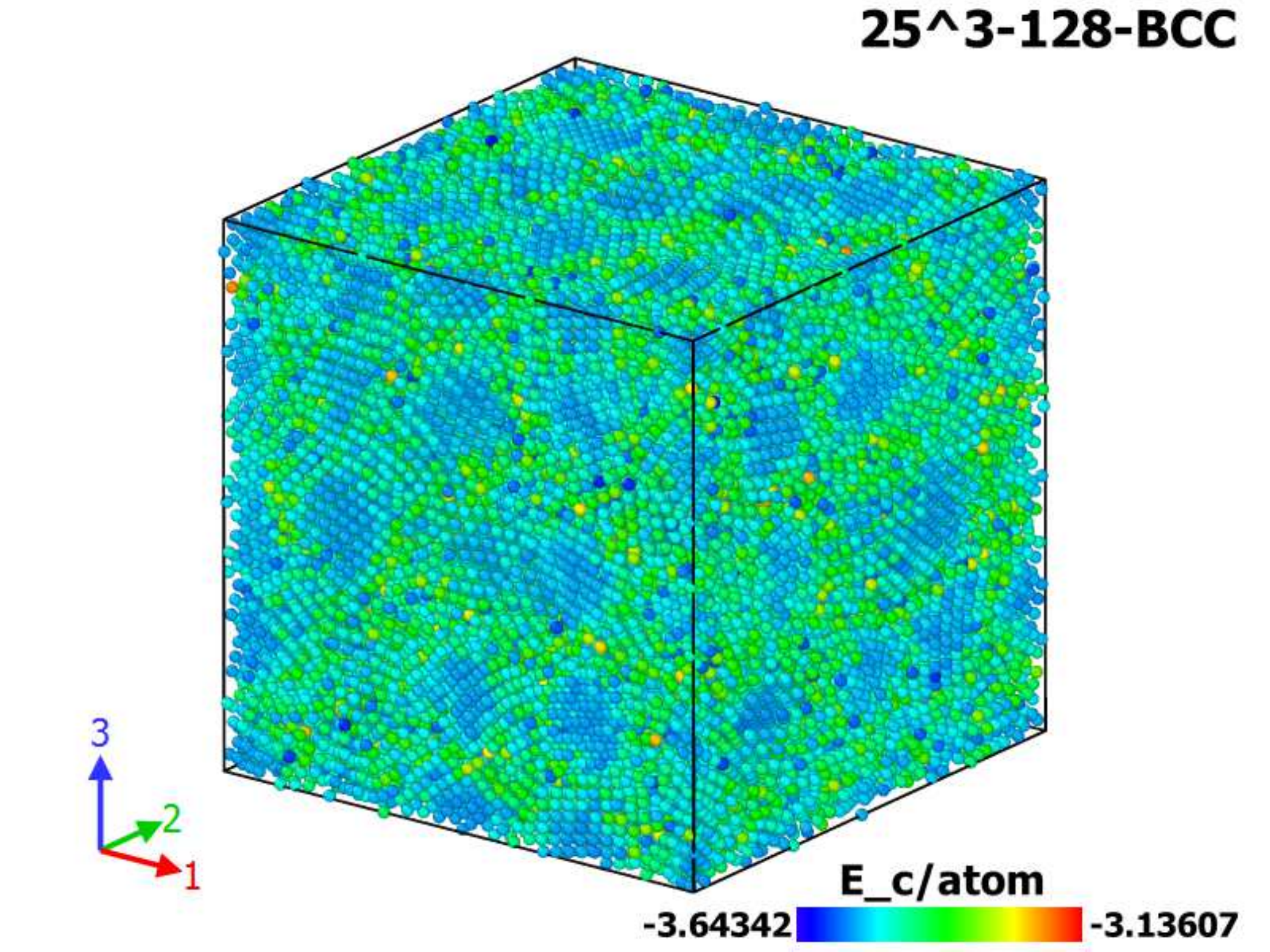} &
		\includegraphics[width=0.34\linewidth]{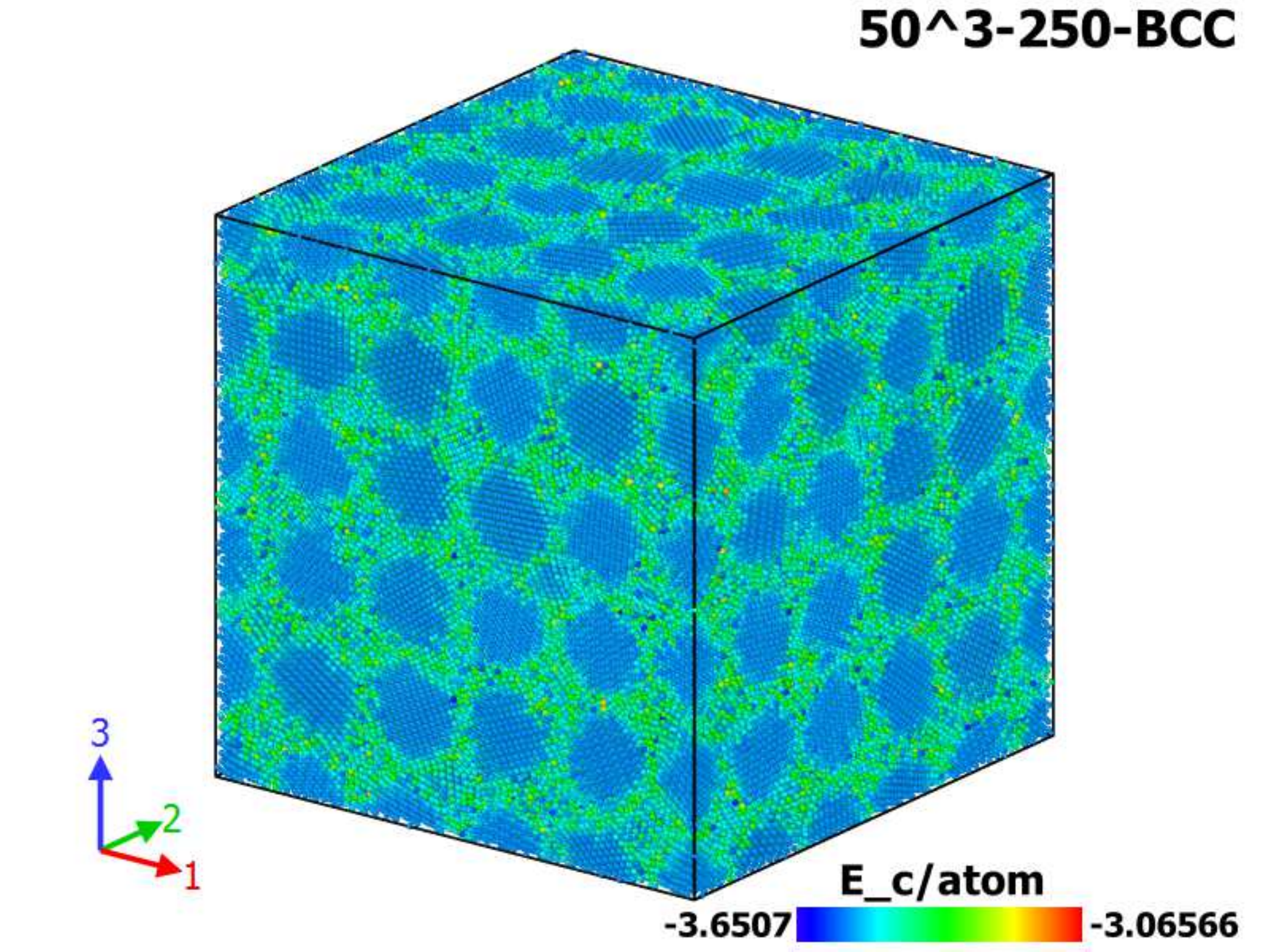} \\
			\includegraphics[width=0.34\linewidth]{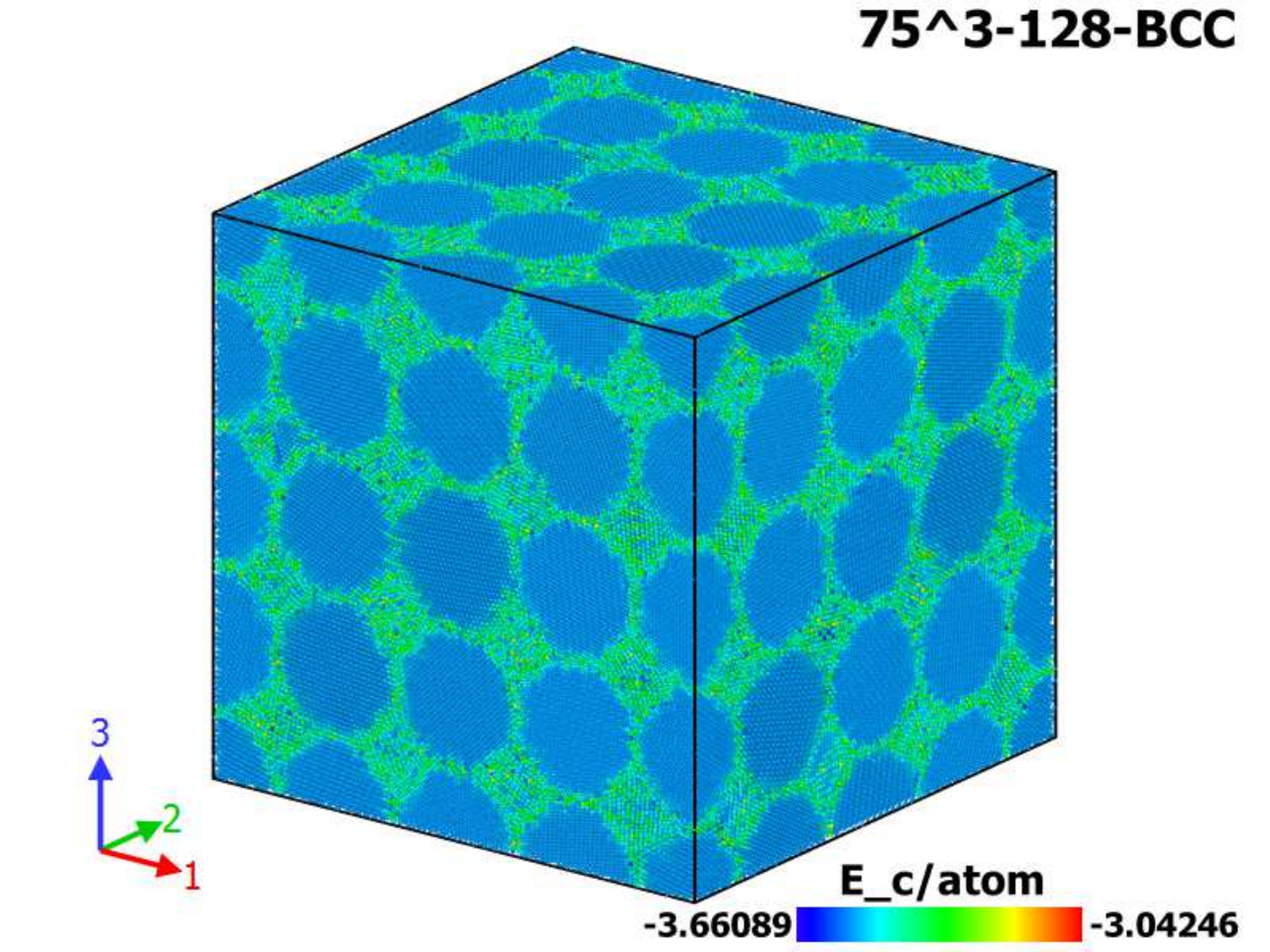} &
		\includegraphics[width=0.34\linewidth]{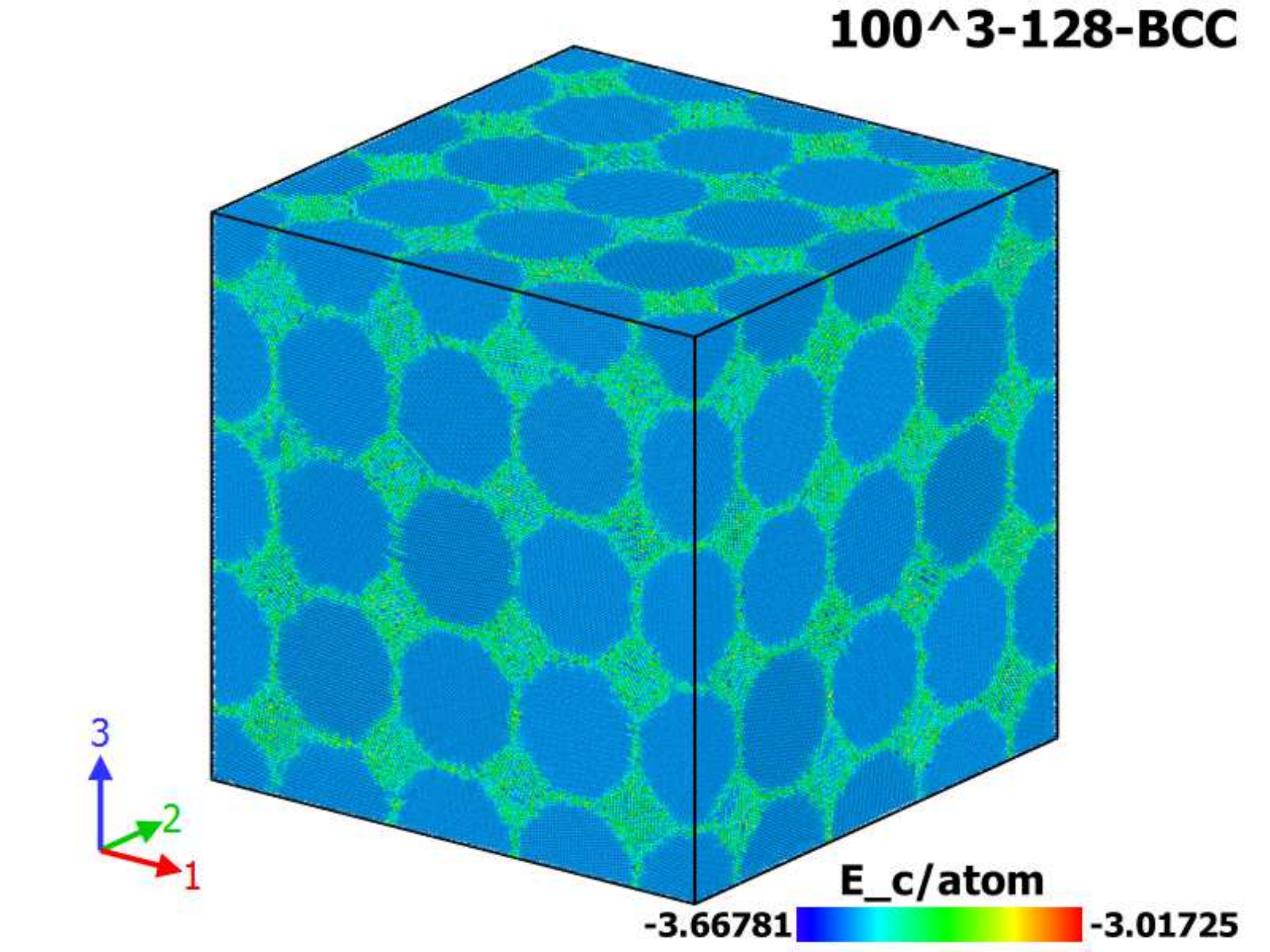} &
		\includegraphics[width=0.34\linewidth]{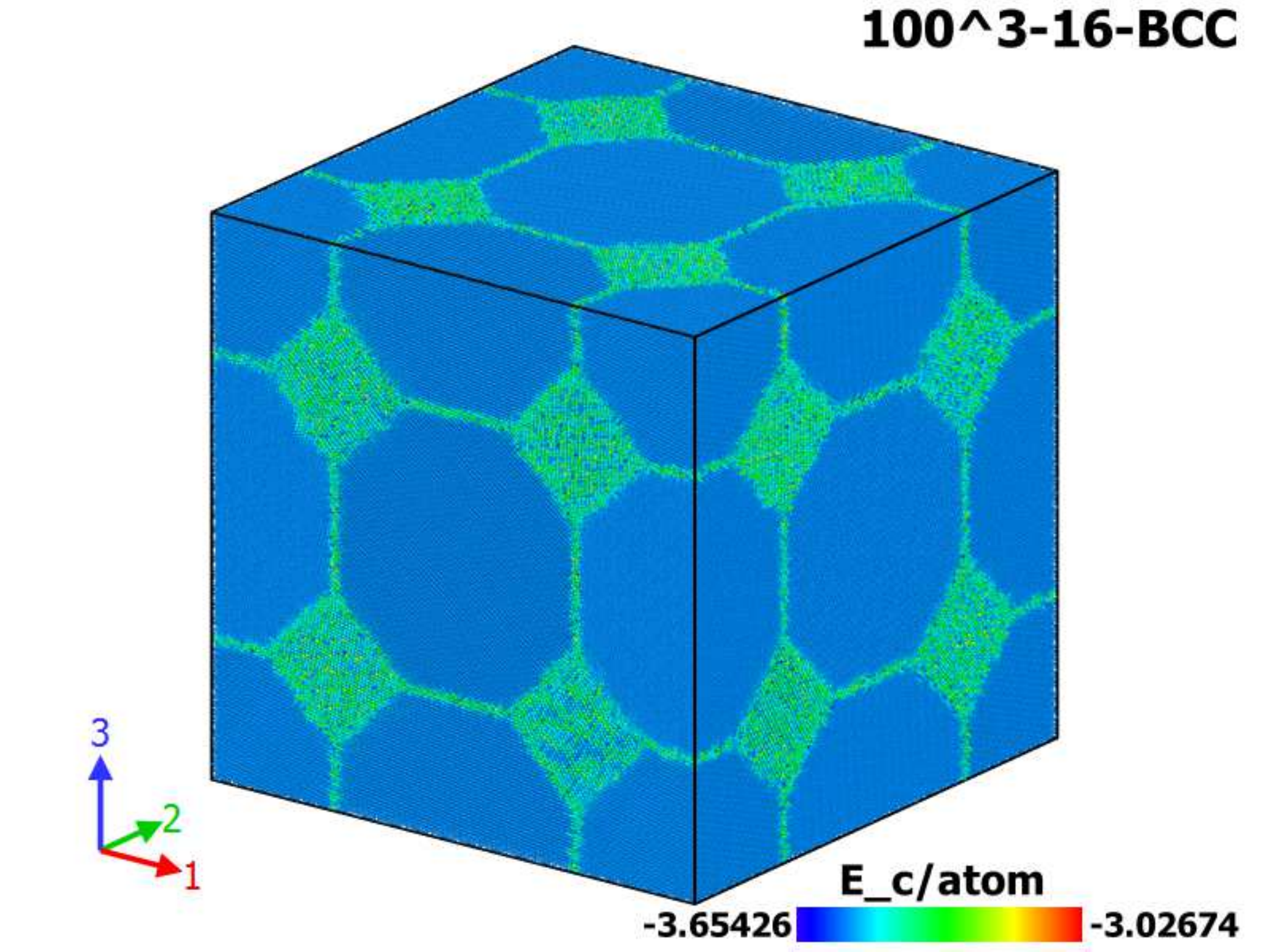} \\
			\includegraphics[width=0.34\linewidth]{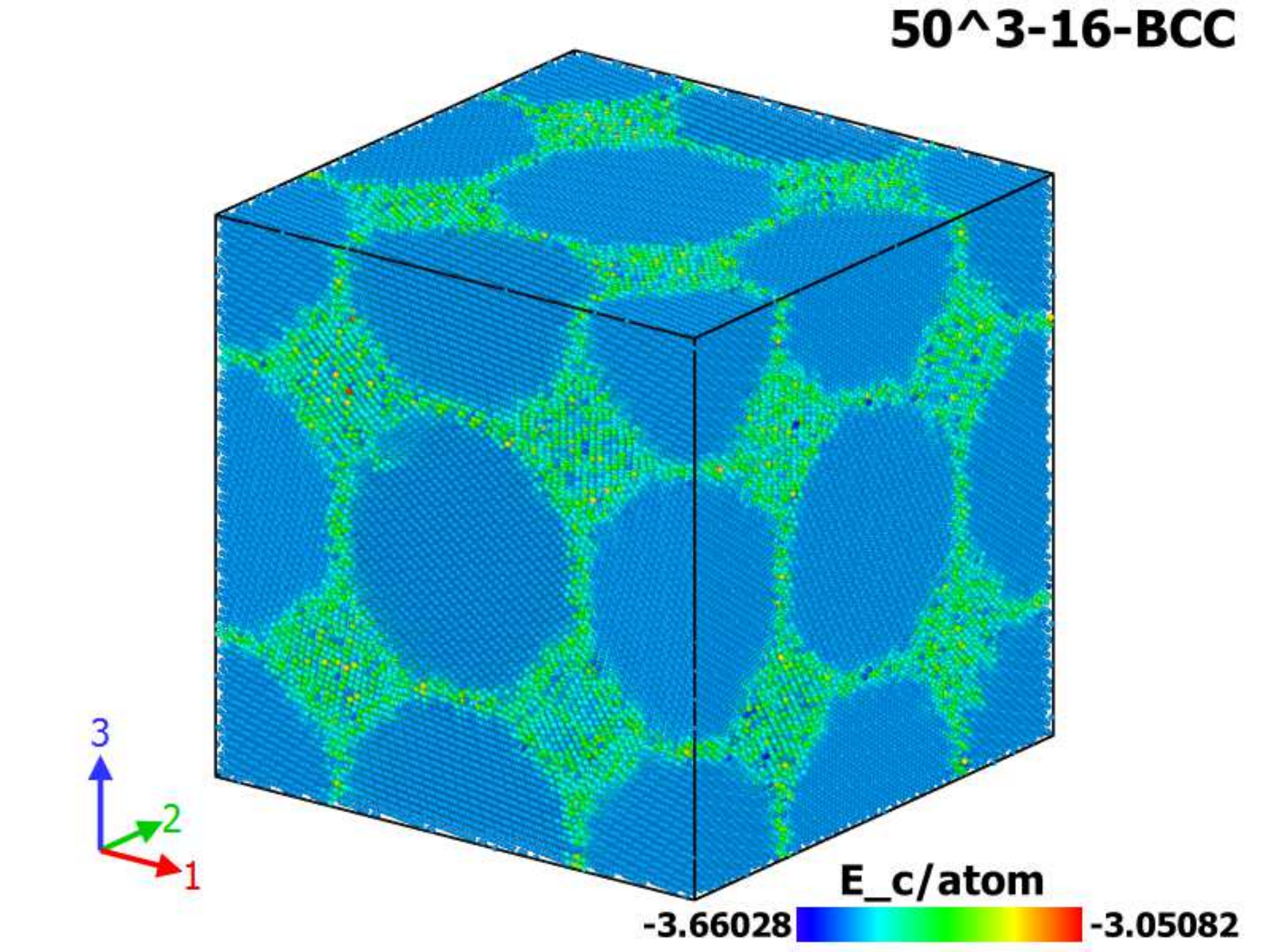} &
		\includegraphics[width=0.34\linewidth]{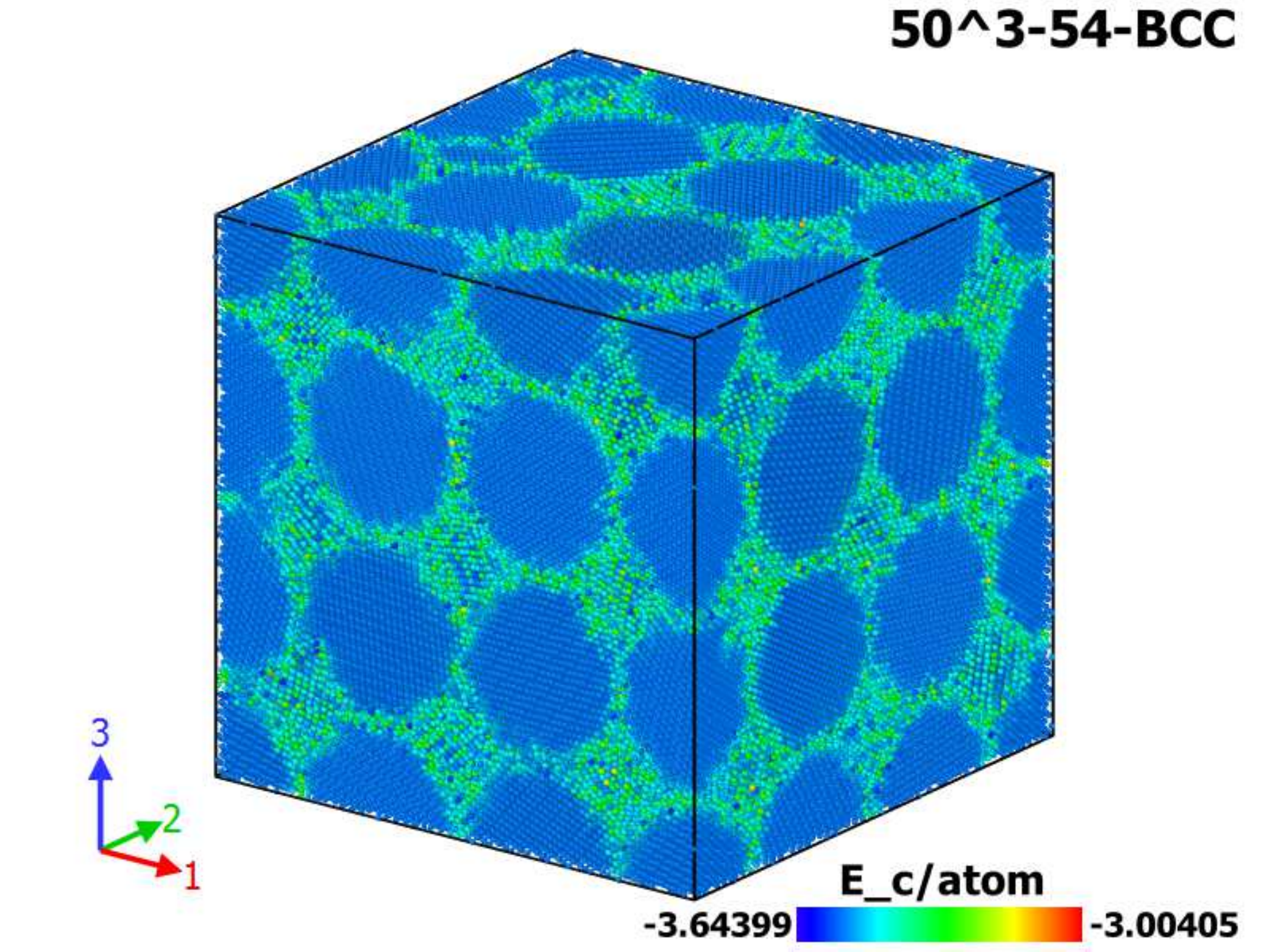} &
		\includegraphics[width=0.34\linewidth]{503-250-BCC-eps-converted-to.pdf} \\
			\includegraphics[width=0.34\linewidth]{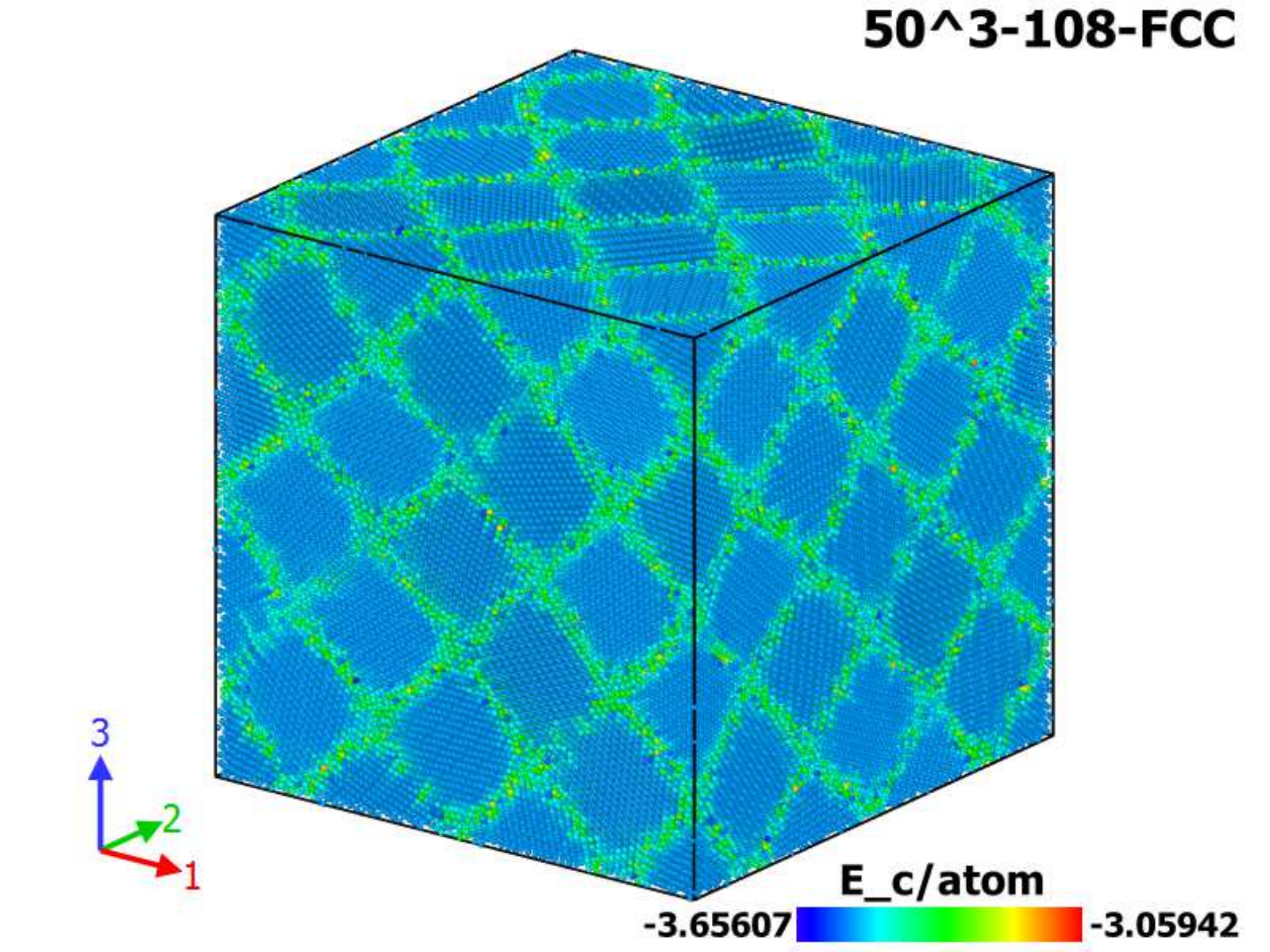} &	
		\includegraphics[width=0.34\linewidth]{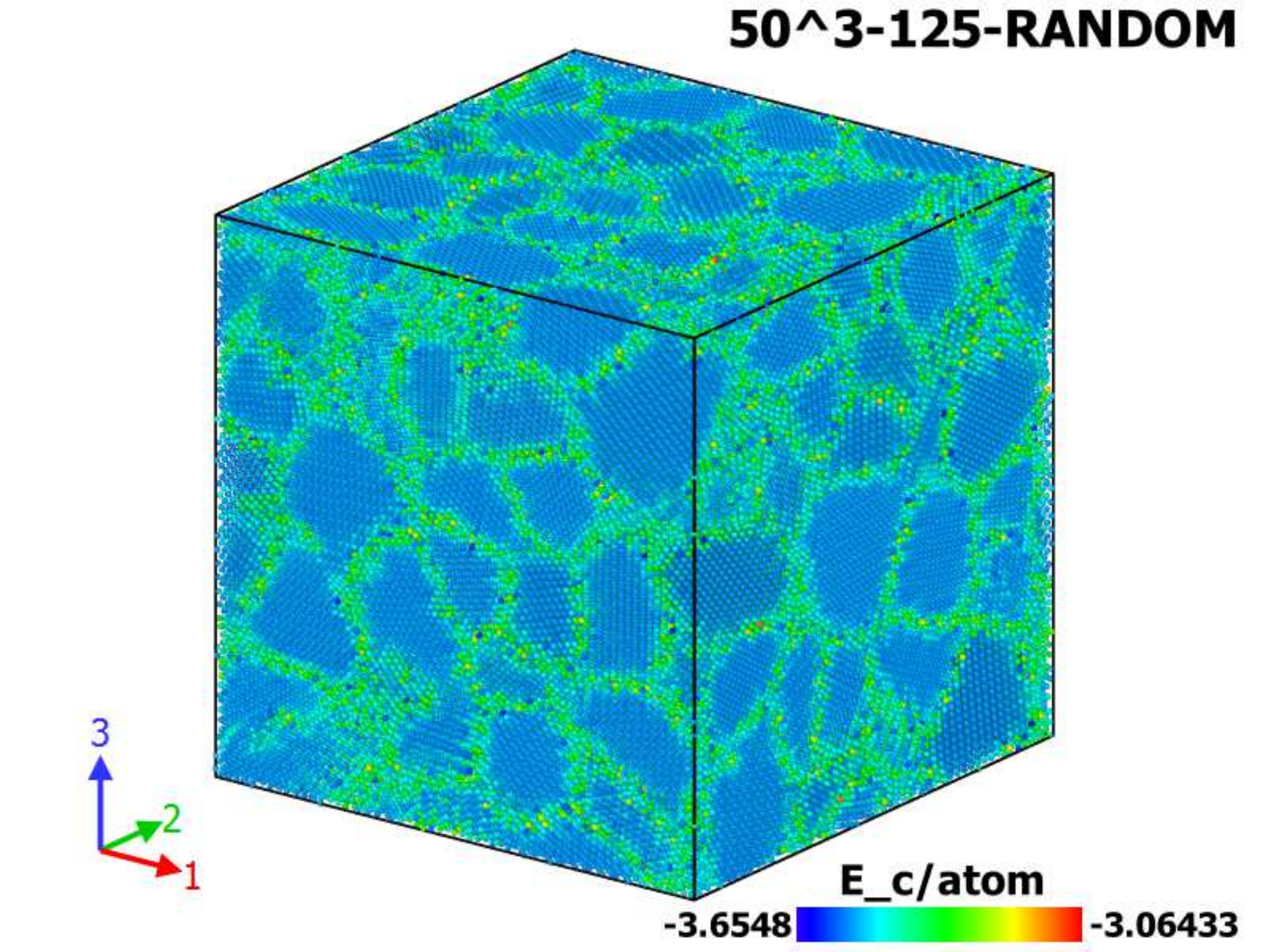}
	\end{tabular}
	\caption{Visualization of atomistic computational samples and cohesive energy $E_{c}$\,(eV/atom).  }
	\label{fig:Samples}
\end{figure}

\begin{figure}[H]
	\centering
	\begin{tabular}{ccc}
		\includegraphics[width=0.34\linewidth]{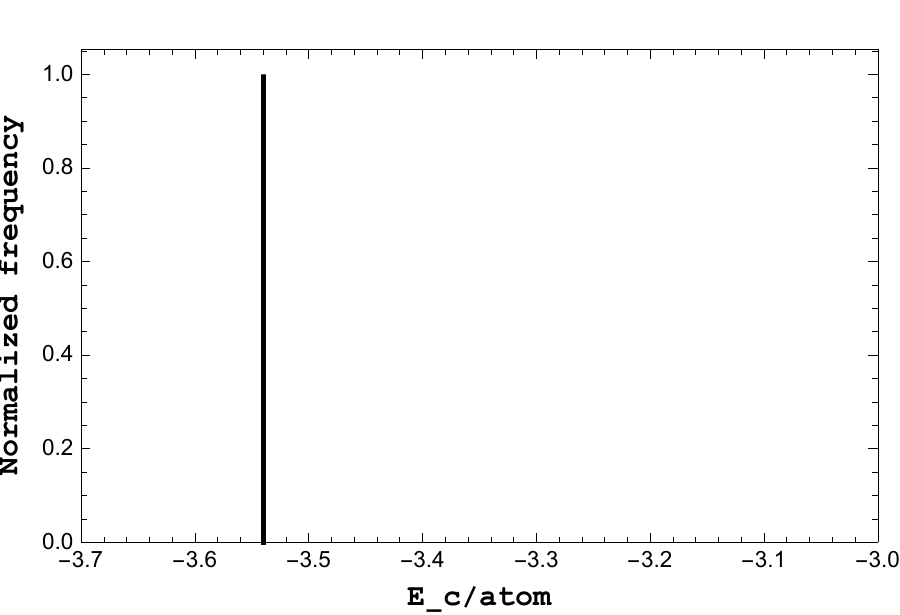} &
		\includegraphics[width=0.34\linewidth]{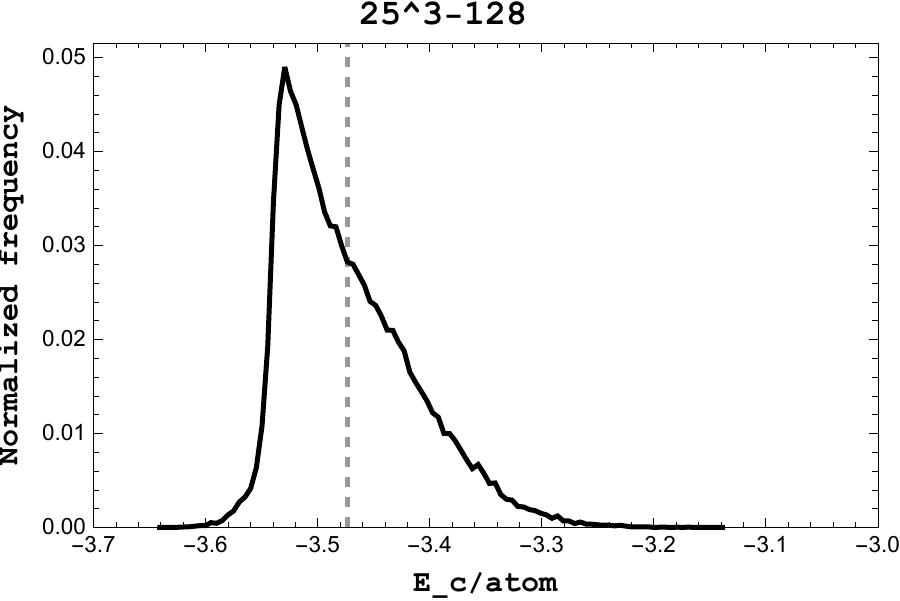} &
		\includegraphics[width=0.34\linewidth]{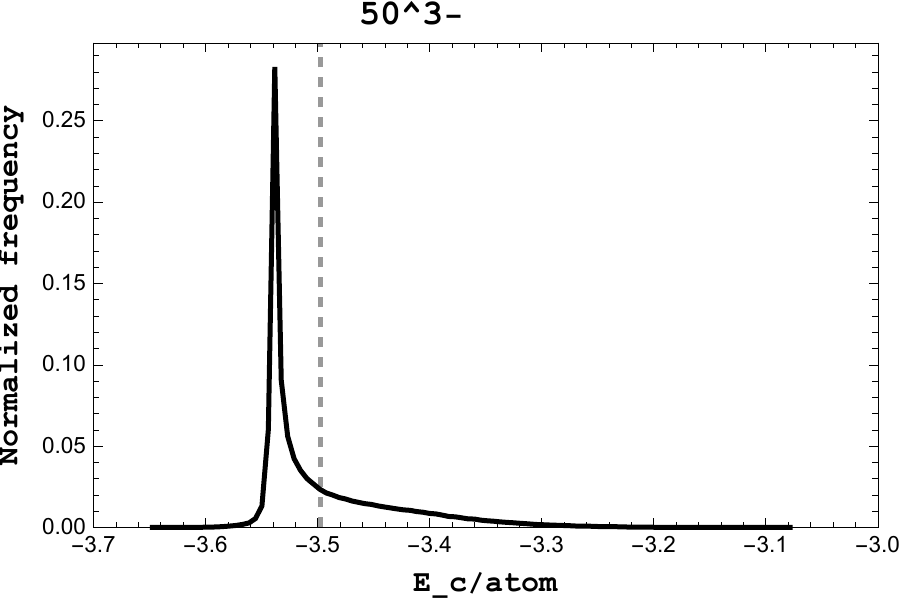} \\
		\includegraphics[width=0.34\linewidth]{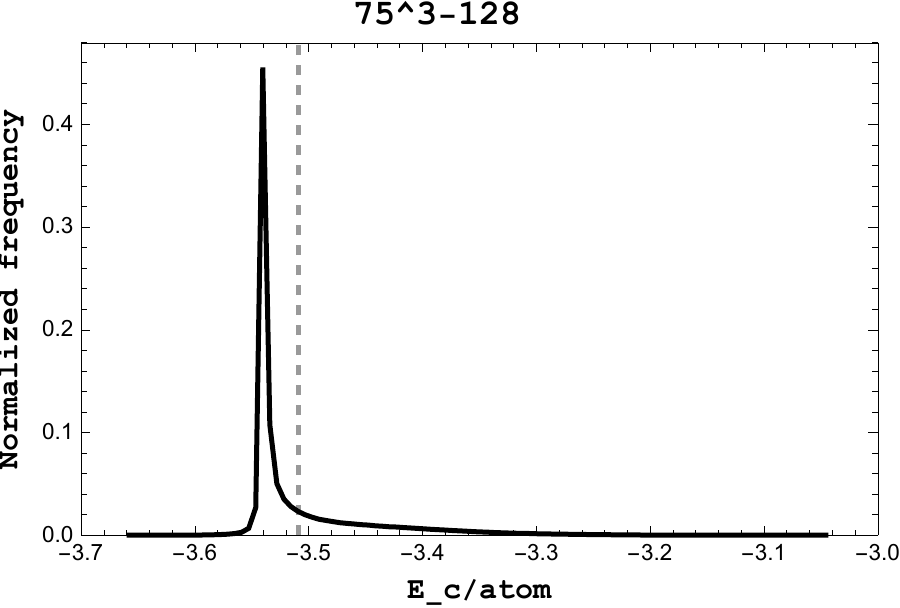} &
		\includegraphics[width=0.34\linewidth]{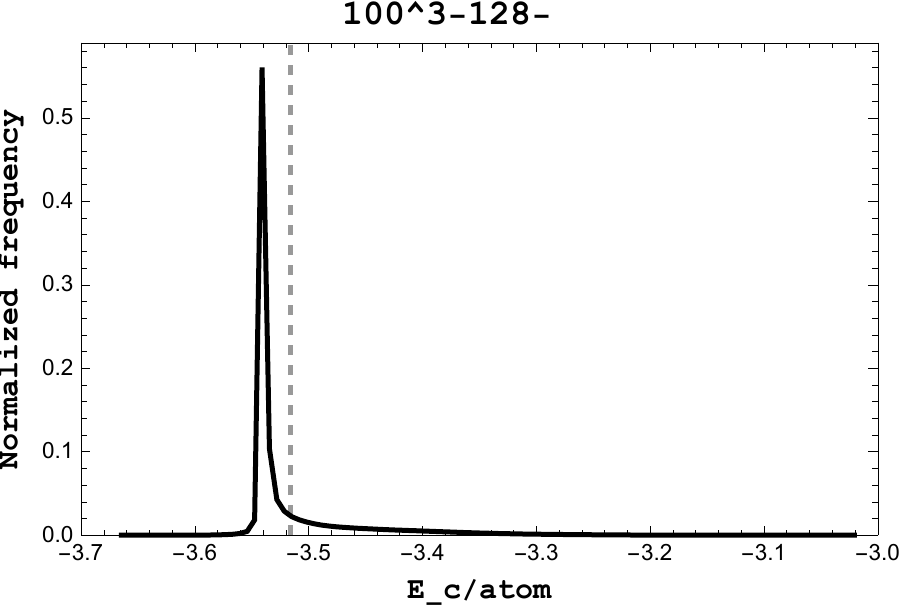} &
		\includegraphics[width=0.34\linewidth]{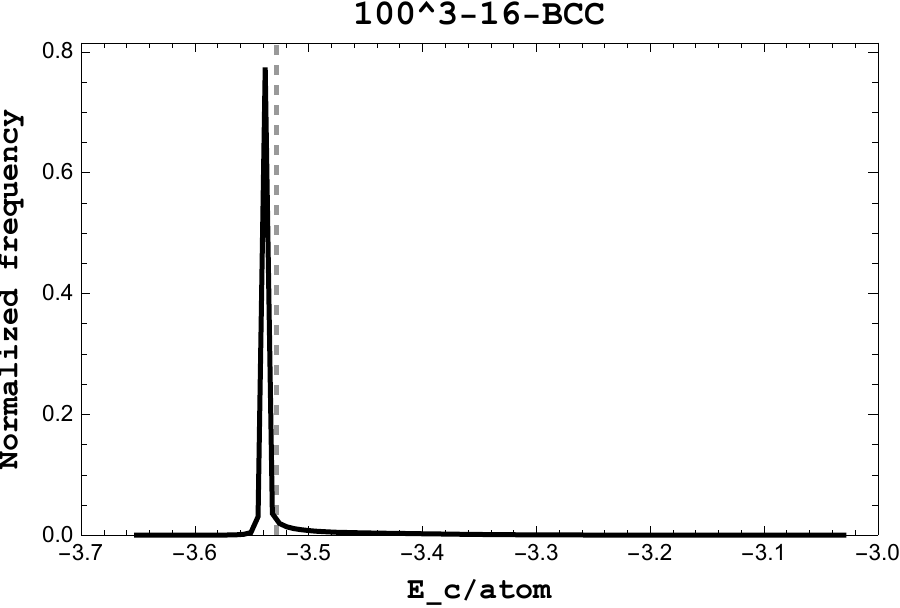} \\
		\includegraphics[width=0.34\linewidth]{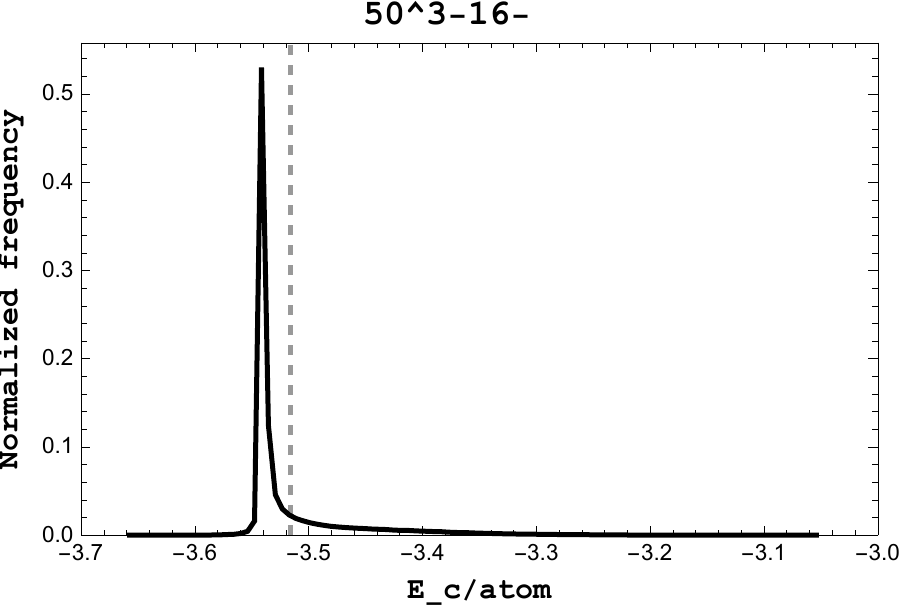} &
		\includegraphics[width=0.34\linewidth]{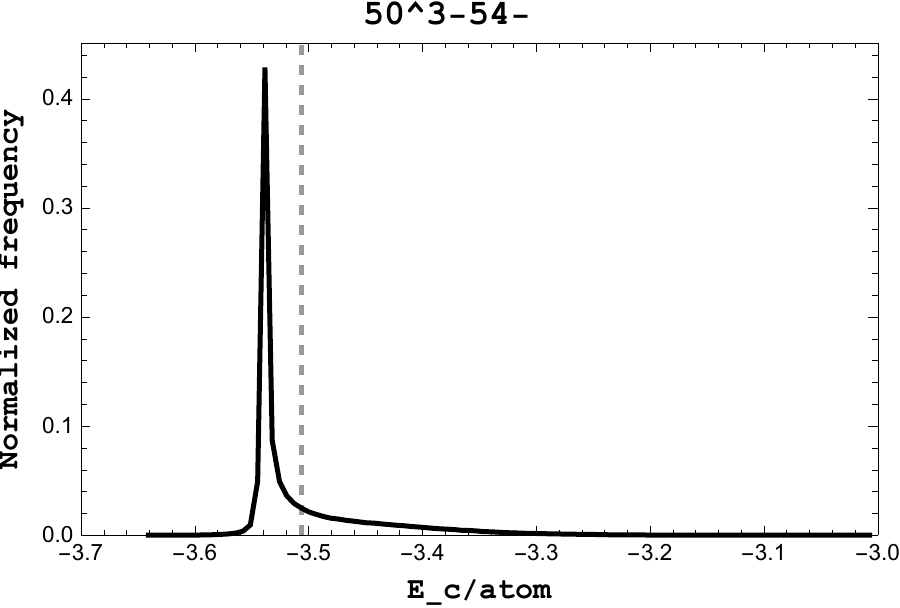} &
		\includegraphics[width=0.34\linewidth]{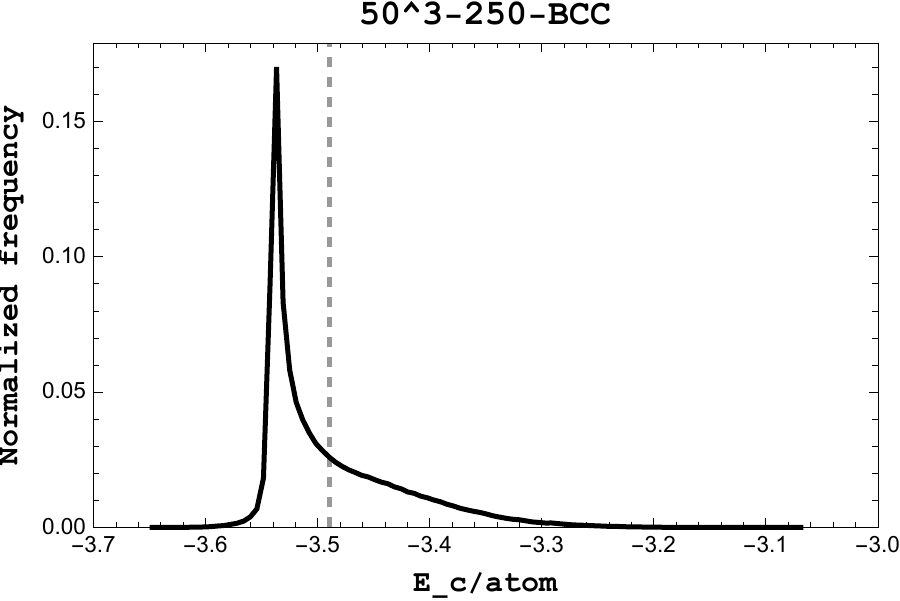} \\
		\includegraphics[width=0.34\linewidth]{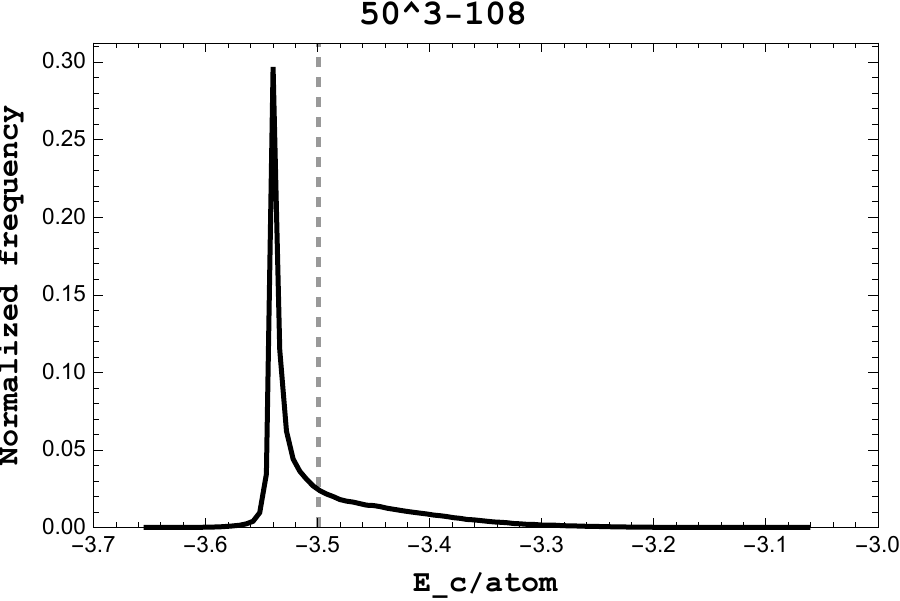} &	
		\includegraphics[width=0.34\linewidth]{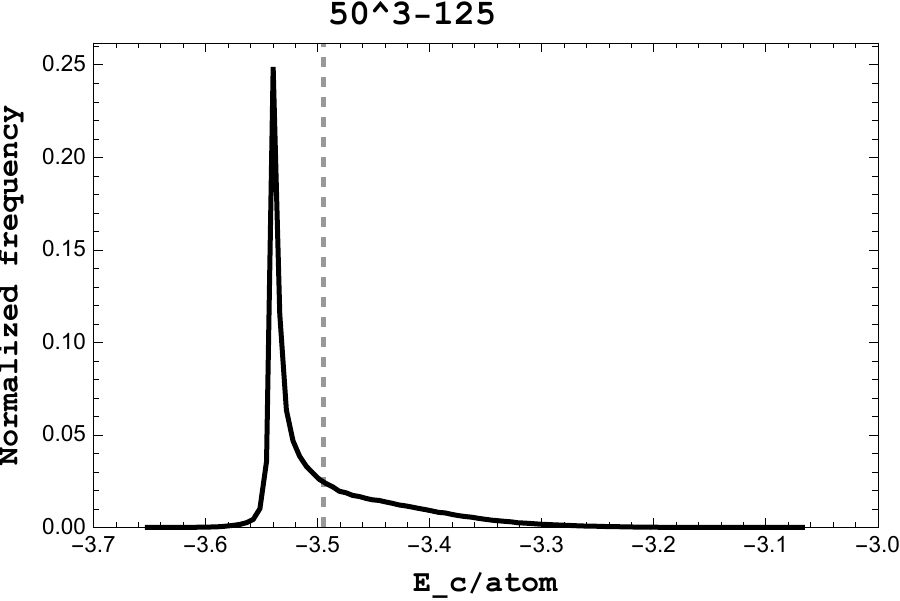} &
		\includegraphics[width=0.34\linewidth]{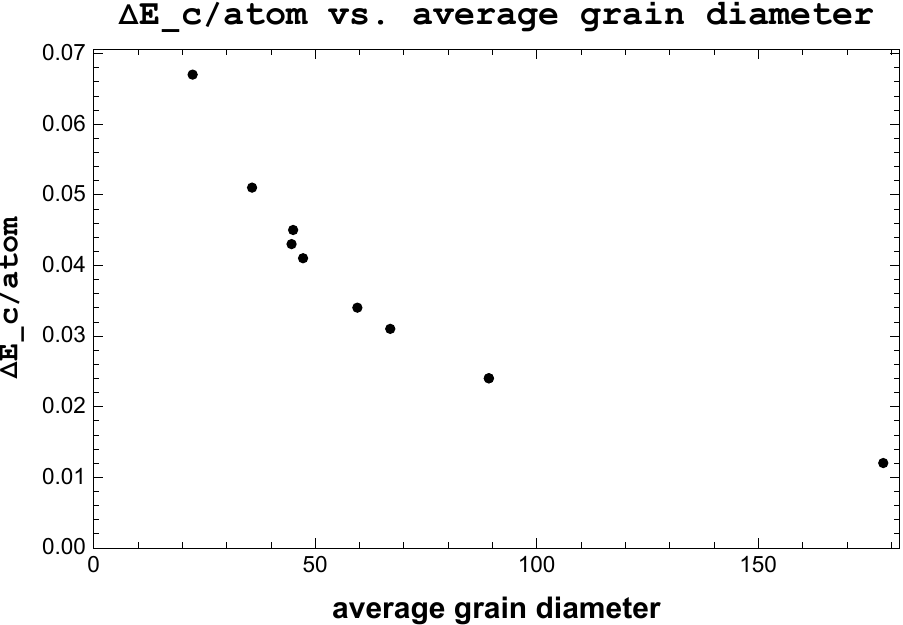}
	\end{tabular}
	\caption{Cohesive energy density $E_{c}$\,(eV/atom) histograms (vertical dashed line represents the average cohesive energy) and $\Delta E_{c}/atom = E_{c}/atom-E^{Monocrystal}_{c}/atom$\,(eV/atom) as a function of average grain diameter $d$ (\AA). }
	\label{fig:Histograms}
\end{figure}

Visualizations of atomistic computational samples and cohesive energy $E_{c}$\,(eV/atom) are depicted in Fig. \ref{fig:Samples}, while cohesive energy density $E_{c}$\,(eV/atom) histograms in Fig. \ref{fig:Histograms}. It can be seen that as the average grain size decreases, the fraction of transient shell atoms in the sample and the average cohesive energy rises.

Fig. \ref{fig:Samples} indicates that polycrystals can be treated as crystalline cores of monocrystal pattern surrounded by amorphous sheaths. In solid-state systems the CNA pattern is a useful measure of the local crystal structure around an atom \cite{Plimpton1995}. There are five kinds of CNA patterns: FCC, HCP, BCC, Icosohedral and Other. Sometimes it may be difficult to choose the right \textit{cutoff} radius for the conventional CNA and therefore an adaptive version of the CNA has been developed that works without a fixed \textit{cutoff}. The adaptive common neighbor analysis (a-CNA) method determines the optimal \textit{cutoff} radius automatically for each individual particle \cite{Stukowski2012}. On the basis of this analysis for FCC copper we can define fraction of non-FCC structure atoms $f_{CNA}$ in the polycrystalline sample, see Table \ref{tab:Samples}. 

In a core-shell model proposed in Sec.\ref{sec:Cont} the grain is composed of a core and of the shell that has different properties (see also \cite{Palosz2002}). In conjunction with atomistic simulations the shell thickness $\Delta$ can be preliminary assessed as equal to \textit{the cutoff radius} of the used potential \cite{Mishin2001}, i.e. 5.5\,\AA.{The value is consistent with the general premises for the non-bulk layers of atoms in a 3D FCC crystals \cite{Park2007}.} Using such definition and assuming spherical shape of coated grains the fraction of transient shell atoms $f_0$ in the sample is calculated by the formula
\begin{equation}\label{def:fsa}
	f_0=1-\left(1-\frac{2\Delta}{d}\right)^3\,,\quad\Delta=5.5\,\text{\AA}
\end{equation} 
where $d$ is an averaged grain diameter. The values of $d$ and $f_0$ for ten analysed samples are collected in Table \ref{tab:Samples}.
 
\begin{table}[H] 
	\caption{Volume (\AA$^3$), box lengths (\AA), number of atoms, average grain diameter $d$ (\AA), fraction of transient shell atoms $f_{0}$ (\ref{def:fsa}), {fraction of non-FCC structure atoms $f_{CNA}$}, average cohesive energy $E_{c}$\,(eV/atom) of analysed computational samples.}
		\label{tab:Samples}
		\centering
		\renewcommand{\arraystretch}{1.5}
		\tiny 
		\begin{tabular}{|c c c c c c c c|}
			\hline Sample & V & L & No.of atoms & $d$ & $f_{0}$ & $f_{CNA}$ & E$_{c}$\\ 
			\hline Monocrystal & 47.24 & 3.615 & 4 & & & & -3.54 \\
			 25$^3$-128-BCC & 750452.29 & 90.87 & 62302 & 22.4 & 0.87 & 0.71 &-3.473 \\
			 50$^3$-128-BCC & 5985342.9 & 181.56 & 499984 & 44.7 & 0.57 & 0.36 &-3.497 \\
			 75$^3$-128-BCC & 20129512 & 272.03 & 1687719 & 67 & 0.42 & 0.25 &-3.509 \\
			 100$^3$-128-BCC & 47605657 & 362.43 & 3999934 & 89.2 & 0.33 & 0.19 & -3.516 \\
			 100$^3$-16-BCC & 47432793 & 361.99 & 4000010 & 178.2 & 0.17 & 0.09 & -3.528 \\
			 50$^3$-16-BCC & 5950371.6 & 181.21 & 500020 &  89.2 & 0.33 & 0.18 & -3.516 \\
			 50$^3$-54-BCC & 5969399.2 & 181.40 & 500058 & 59.6 & 0.46 & 0.28 &-3.506 \\
			 50$^3$-250-BCC & 5997791.1 & 181.69 & 500008 & 35.8 & 0.67 & 0.46 &-3.489 \\
			 50$^3$-108-FCC & 5981645.7 & 181.53 & 500021 &  47.3 & 0.55 & 0.35 &-3.499 \\
			 50$^3$-125-Random & 5985474.7 & 181.57 & 499836 & 45.1 & 0.57 & 0.39 &-3.495 \\
			\hline 
		\end{tabular}
\end{table}

\begin{table}[H] 
	\caption{{{Elasticity tensors $\bar{\mathbb{C}}$\,[GPa]  of analysed samples (for notation used see Eq. (\ref{eqn:CuCij}) in appendix)}}.}
	\label{tab:Cij}
	\begin{threeparttable}[b]
		\centering
		\tiny
\begin{tabular}{ c c }
	{Monocrystal} & {25$^3$ unit cells, 128 grains in BCC system, 62302 atoms} \\
		 & {(25$^3$-128-BCC)} \\
	$\begin{bmatrix}  {169.88} & {122.60} & {122.60} & 0 & 0 & 0 \\
	{122.60} & {169.88} & {122.60} & 0 & 0 & 0 \\
	{122.60} & {122.60} & {169.88} & 0 & 0 & 0 \\
	0 & 0 & 0 & {76.19} & 0 & 0 \\
	0 & 0 & 0 & 0 & {76.19} & 0 \\
	0 & 0 & 0 & 0 & 0 & {76.19} \end{bmatrix}$ &
		
$\begin{bmatrix}  {166.50} & {120.68} & {120.46} & {-0.56} & {0.53} & {0.65} \\
{120.68} & {167.56} & {120.64} & {-1.64} & {-0.15} & {0.76} \\
{120.46} & {120.64} & {166.60} & {3.42} & {-0.57} & {-0.32} \\
{-0.56} & {-1.64} & {3.42} & {22.62} & {0.06} & {-0.39} \\
{0.53} & {-0.15} & {-0.57} & {0.06} & {23.96} & {-0.56} \\
{0.65} & {0.76} & {-0.32} & {-0.39} & {-0.56} & {22.09} \end{bmatrix}$\\  

 \hline {50$^3$ unit cells, 128 grains in BCC system, 499984 atoms} & {75$^3$ unit cells, 128 grains in BCC system, 1687719 atoms} \\  
  {(50$^3$-128-BCC)} & {(75$^3$-128-BCC)} \\  
	$\begin{bmatrix}  {174.61} & {116.92} & {119.13} & {-0.34} & {1.11} & {-1.54} \\
	{116.92} & {177.10} & {117.18} & {-1.80} & {1.09} & {1.17} \\
	{119.13} & {117.18} & {174.8778} & {0.18} & {-0.27} & {0.08} \\
	{-0.34} & {-1.80} & {0.18} & {32.96} & {0.17} & {-1.63} \\
	{1.11} & {1.09} & {-0.27} & {0.17} & {28.10} & {1.05} \\
	{-1.54} & {1.17} & {0.08} & {-1.63} & {1.05} & {29.19} \end{bmatrix}$ &
	
	$\begin{bmatrix}  {180.71} & {114.88} & {115.79} & {-1.074} & {-1.39} & {-0.57} \\
	{114.88} & {181.95} & {115.82} & {-0.94} & {-0.62} & {1.40} \\
	{115.79} & {115.82} & {180.43} & {1.58} & {0.64} & {-0.85} \\
	{-1.074} & {-0.94} & {1.58} & {34.986} & {-0.44} & {-0.07} \\
	{-1.39} & {-0.62} & {0.64} &  {-0.44} & {36.91} & {0.39} \\
	{-0.57} & {1.40} & {-0.85} & {-0.07} & {0.39} & {35.74} \end{bmatrix}$\\

\hline {100$^3$ unit cells, 128 grains in BCC system, 3999934 atoms} & {100$^3$ unit cells, 16 grains in BCC system, 4000010 atoms} \\ 
{(100$^3$-128-BCC)} & {(100$^3$-16-BCC)} \\ 	
	$\begin{bmatrix}  {184.74} & {113.60} & {115.07} & {-0.61} & {0.34} & {-0.87} \\
	{113.60} & {184.55} & {114.18} & {-0.69} & {-0.92} & {1.83} \\
	{115.07} & {114.18} & {183.48} & {1.54} & {0.22} & {-0.82} \\
	{-0.61} & {-0.69} & {1.54} & {37.90} & {-0.43} &  {-0.44} \\
	{0.34} & {-0.92} & {0.22} & {-0.43} & {37.57} & {-0.57}\\
	{-0.87} & {1.83} & {-0.82} & {-0.44} & {-0.57} & {37.93} \end{bmatrix}$ &
	
	$\begin{bmatrix}  {186.97} & {114.79} & {112.47} & {-0.88} & {-0.07} & {-6.21} \\
	{114.79} & {186.11} & {113.14} & {-1.20} & {1.42} & {2.40} \\
	{112.47} & {113.14} & {187.21} & {2.26} & {-1.08} & {3.56} \\
	{-0.88} & {-1.20} & {2.26} & {42.93} & {2.71} & {1.77} \\
	{-0.07} & {1.42} & {-1.08} & {2.71} & {42.50} & {-1.00} \\
	{-6.21} & {2.40} & {3.56} & {1.77} & {-1.00} & {44.38} \end{bmatrix}$\\
	
\hline {50$^3$ unit cells, 16 grains in BCC system, 500020 atoms} & {50$^3$ unit cells, 54 grains in BCC system, 500058 atoms} \\
{(50$^3$-16-BCC)} & {(50$^3$-54-BCC)} \\
	$\begin{bmatrix}  {182.57} & {115.48} & {115.55} & {-0.65} & {0.03} & {-6.23} \\
	{115.48} & {182.04} & {115.56} & {-1.44} & {0.09} & {3.82} \\
	{115.55} & {115.56} & {180.12} & {1.32} & {-0.69} & {2.28} \\
	{-0.65} & {-1.44} & {1.32} & {38.88} & {3.44} & {0.88} \\
	{0.03} & {0.09} & {-0.69} & {3.44} & {38.67} & {-1.52} \\
	{-6.23} & {3.82} & {2.28} & {0.88} & {-1.52} & {40.46} \end{bmatrix}$ &

	$\begin{bmatrix}  {177.33} & {116.96} & {114.38} & {-1.49} & {-0.49} & {-2.32} \\
	{116.96} & {179.06} & {116.59} & {-1.01} & {-1.65} & {2.03} \\
	{114.38} & {116.59} & {175.49} & {3.30} & {3.30} & {0.30} \\
	{-1.49} & {-1.01} & {3.30} & {36.85} & {1.42} & {-0.36} \\
	{-0.49} & {-1.65} & {3.30} & {1.42} & {36.20} &  {-1.05}\\
	{-2.32} & {2.03} & {0.30} & {-0.36} & {-1.05} & {35.09} \end{bmatrix}$\\

\hline {50$^3$ unit cells, 250 grains in BCC system, 500008 atoms} & {50$^3$ unit cells, 108 grains in FCC system, 500021 atoms} \\
{(50$^3$-250-BCC)} & {(50$^3$-108-FCC)} \\
$\begin{bmatrix}  {172.29} & {118.51} & {120.44} & {-0.14} & {0.45} & {-1.32} \\
{118.51} & {173.39} & {119.08} & {-0.48} & {0.46} & {0.35} \\
{120.44} & {119.08} & {173.41} & {1.24} & {-0.55} & {0.75} \\
{-0.14} & {-0.48} & {1.24} & {28.53} & {-2.40} & {-1.77} \\
{0.45} & {0.46} & {-0.55} & {-2.40} & {27.57} & {-1.19} \\
{-1.32} & {0.35} & {0.75} & {-1.77} & {-1.19} & {29.23} \end{bmatrix}$ &

$\begin{bmatrix}  {177.94} & {117.26} & {116.47} & {-1.255} & {0.43} & {1.01} \\
{117.26} & {174.99} & {119.01} & {-0.92} & {-1.33} & {1.377} \\
{116.47} & {119.01} & {175.09} & {1.67} & {1.60} & {-0.49} \\
{-1.255} & {-0.92} & {1.67} & {33.06} & {-0.14} & {0.46} \\
{0.43} & {-1.33} & {1.60} & {-0.14} & {29.84} & {-0.75} \\
{1.01} & {1.377} & {-0.49} & {0.46} & {-0.75} & {33.20} \end{bmatrix}$\\

\hline \multicolumn{2}{ c }{50$^3$ unit cells, 125 grains in random system, 500008 atoms} \\
 \multicolumn{2}{ c }{(50$^3$-125-Random)} \\
	\multicolumn{2}{ c }{$\begin{bmatrix}  {176.96} & {115.67} & {118.06} & {3.00} & {2.65} & {-2.05} \\
	{115.67} & {174.31} & {118.15} & {-0.71} & {-1.01} & {-0.10} \\
	{118.06} & {118.15} & {175.02} & {3.97} & {1.02} & {0.04} \\
	{3.00} &  {-0.71}  & {3.97} & {31.71} & {2.56} & {-2.30} \\
	{2.65} & {-1.01} & {1.02} & {2.56} & {30.66} & {0.05} \\
	{-2.05} & {-0.10} & {0.04} & {-2.30} & {0.05} & {29.92} \end{bmatrix}$ }\\
\end{tabular}
	\end{threeparttable}
    \end{table}

\subsection{Comparison of atomistic and continuum-mechanics estimates}
\label{ssec:CompACest}

Table~\ref{tab:BoundsCM} contains classical continuum-mechanics mean-field estimates (\ref{lowerL})-(\ref{sc1}) of the effective isotropic shear modulus (\ref{Eq:IsoBulkShear})$_2$ and anisotropy factors (\ref{Eq:zeta1}-\ref{Eq:zeta2}) calculated  for the respective sets of $N_g$ orientations assumed in the atomistic simulations. As discussed in \ref{Ap:A}, such estimates neglect the grain size effect and treat the polycrystal as a one-phase material. Table~\ref{tab:BoundsCM} demonstrates that the mean-field  estimates $\bar{G}^{\mathcal{L}}_{\rm{iso}}$ are close to the corresponding estimates obtained for the infinite set of orientations. When the number of orientations increases anisotropy factors approach limit values characterizing isotropy. Moreover, for a given $N_g$ the factors are close to each other for different estimates. 
Therefore, in further analysis of atomistic predictions the estimates of the bulk and shear moduli obtained for an infinite set of orientations will be used as reference values.

For comparison purpose, the anisotropy degree of effective stiffness $\bar{\mathbb{C}}$, obtained for a given sample in atomistic simulations were also analysed (see Table \ref{tab:AtomEstimates}).     
It was observed that anisotropy of effective properties acquired in atomistic simulations, when assessed using the anisotropy factors (\ref{Eq:zeta1}) and (\ref{Eq:zeta2}), is higher than the corresponding values obtained for the same number of orientations $N_g$ employing the continuum mechanics (see Table \ref{tab:BoundsCM}). Additionally,  for the same number of orientations anisotropy of atomistic estimates decreases with increasing size of the grain specified by the number of atoms per grain. On the other hand, for the same number of unit cells $N_{UC}$  (i.e. the similar number of atoms per sample) anisotropy of atomistic estimates decreases with an increasing number of orientations.

\begin{table}[!htp]
	\caption{The overall shear modulus $\bar{G}^{\mathcal{L}}_{\rm{iso}}$\,[GPa] and anisotropy factors $\zeta_1$ and $\zeta_2$\,[\%] obtained for the Voigt, Reuss and Hashin-Shtrikman (H-S) bounds of effective stiffness tensor for copper polycrystal composed of $N_g$ grains. The local elastic stiffness (\ref{Eq:SpekCubic}) is specified by: $K=138.36$\,GPa, $G_1=23.64$\,GPa, $G_2=76.19$\,GPa.
	}
	\label{tab:BoundsCM}\vspace{.05in}
	\centering
	\begin{tabular}{|c|cccccc|cccccc|}
		\hline
		$N_g$ & $\bar{G}^{\mathcal{L}}_{\rm{iso}}$ & $\zeta_1$ & $\zeta_2$ & $\bar{G}^{\mathcal{L}}_{\rm{iso}}$ &  $\zeta_1$ & $\zeta_2$& $\bar{G}^{\mathcal{L}}_{\rm{iso}}$ & $\zeta_1$ & $\zeta_2$ & $\bar{G}^{\mathcal{L}}_{\rm{iso}}$ &  $\zeta_1$ & $\zeta_2$\\ \hline
		1& 47.71 & 0.31 & 10.76 & -- & -- & --& -- & -- & -- & -- & -- & --\\ \hline
		&\multicolumn{3}{c}{Voigt}& \multicolumn{3}{c|}{Reuss}&\multicolumn{3}{c}{H-S (U)}& \multicolumn{3}{c|}{H-S (L)}\\
		16& 54.98 & 0.77 & 1.54 & 40.54 & 0.73 & 1.98& 49.84 & 0.75 & 1.79 & 46.39 & 0.73 & 1.91\\
		54& 55.06 & 0.83 & 1.15 & 40.45 & 0.80 & 1.47& 49.87 & 0.81 & 1.34 & 46.37 & 0.80 & 1.42\\
		128& 55.15 & 0.93 & 0.49 & 40.35 & 0.91 & 0.63& 49.89 & 0.91 & 0.57 & 46.35 & 0.91 & 0.61\\
		250& 55.16 & 0.94 & 0.39 & 40.34 & 0.93 & 0.50 & 49.89 & 0.93 & 0.45 & 46.35 & 0.93 & 0.49\\ \hline
		$\infty$ & 55.17 & 1 & 0 & 40.33 & 1 & 0 & 49.90 & 1 & 0 & 46.35 & 1 & 0 \\ \hline
		
	\end{tabular}\\[.051in]
\end{table}

\begin{table}[!htp]
	\caption{The overall bulk and shear moduli $\bar{K}^{\mathcal{L}}_{\rm{iso}}$\,[GPa] and $\bar{G}^{\mathcal{L}}_{\rm{iso}}$\,[GPa] and anisotropy factors $\zeta_1$ and $\zeta_2$\,[\%] calculated for the effective stiffness tensors resulting from the atomistic simulations and from the MT and SC core-shell models for copper polycrystal ($\Delta$=5.5\,\AA). 
	}
	\label{tab:AtomEstimates}\vspace{.05in}
	\centering
	\begin{tabular}{|c|cccc|ccc|ccc|}
		\hline
	
		Sample & $\bar{K}^{\mathcal{L}}_{\rm{iso}}$ & $\bar{G}^{\mathcal{L}}_{\rm{iso}}$ & $\zeta_1$ & $\zeta_2$  & $\bar{G}^{\mathcal{L}}_{\rm{iso}}$ & $\zeta_1$ & $\zeta_2$& $\bar{G}^{\mathcal{L}}_{\rm{iso}}$ & $\zeta_1$ & $\zeta_2$\\ \hline
			&\multicolumn{4}{c|}{Atomistic}&\multicolumn{3}{c|}{MT Core-Shell}&\multicolumn{3}{c|}{SC Core-Shell}\\ \hline
		$25^3$-128-BCC& 136.02 & 22.91  & 0.78 & 1.80 &25.80& 0.99 & 0.10&25.95&0.99&0.11\\
		$50^3$-128-BCC& 137.00 & 29.50 & 0.78 & 1.71 &31.52& 0.96& 0.32&32.45&0.96&0.34\\
		$75^3$-128-BCC& 137.34 & 34.56 & 0.84 & 1.19 &34.84& 0.95& 0.42&36.35&0.95&0.44\\
		$100^3$-128-BCC& 137.61 & 36.62 & 0.86 & 1.10 &37.00& 0.94& 0.48&38.91&0.94&0.49\\ \hline
		$50^3$-16-BCC& 137.54 & 36.43 & 0.69 & 2.86 & 37.01& 0.82& 1.51&38.90&0.81&1.55\\ 
		$100^3$-16-BCC & 137.90 & 40.26 & 0.70 & 2.59  &41.25& 0.78& 1.85&43.82&0.78&1.83\\ 
		$50^3$-54-BCC & 136.41 & 33.65 & 0.73 & 2.38 & 33.92& 0.89& 0.92&35.26&0.89&0.97\\
		$50^3$-250-BCC & 137.24 & 27.69 & 0.79 & 1.84   & 29.49& 0.98& 0.20&30.09&0.97&0.21\\ \hline
		$50^3$-108-FCC & 137.05 & 30.79 & 0.79 & 1.74 & 31.94& 0.95& 0.43&32.94&0.95&0.45\\
		$50^3$-125-RAN & 136.63 & 29.91 & 0.73 & 2.16   & 31.52& 0.96& 0.32&32.45&0.96&0.34\\ \hline
	\end{tabular}\\[.051in]
\end{table}

\begin{figure}[H]
	\centering
	\includegraphics[width=.8\textwidth]{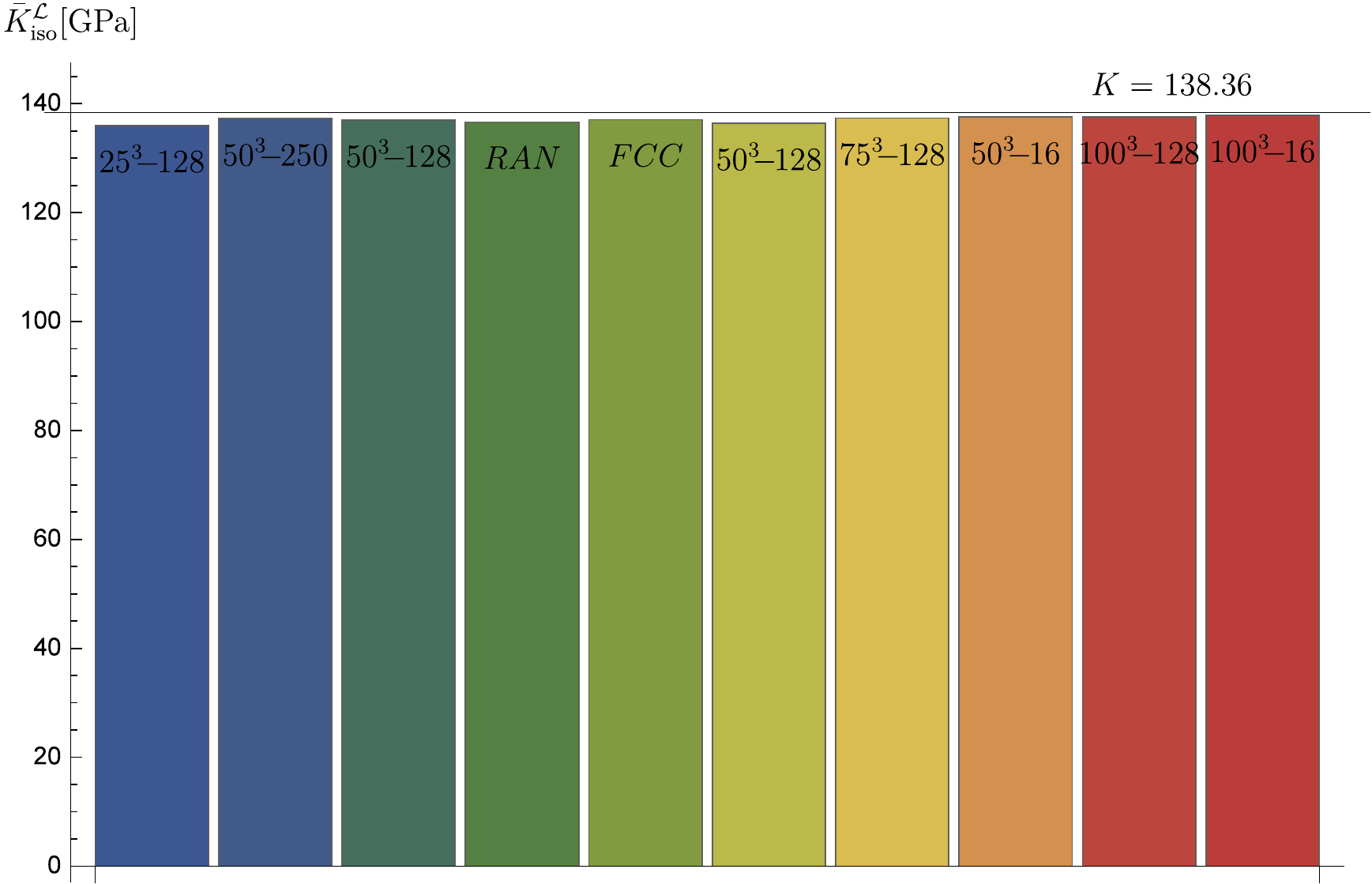}
	\caption{The isotropic bulk modulus $\bar{K}^{\mathcal{L}}_{\rm{iso}}$ calculated for the overall stiffness tensor acquired in atomistic simulations for different polycrystalline samples with respect to the $\bar{K}$ value predicted by the continuum mechanics methodology. Samples are ordered according to the increasing value of grain diameter $d$.}
	\label{fig:BulkA}
\end{figure}

\begin{figure}[H]
	\centering
	\includegraphics[width=.93\textwidth]{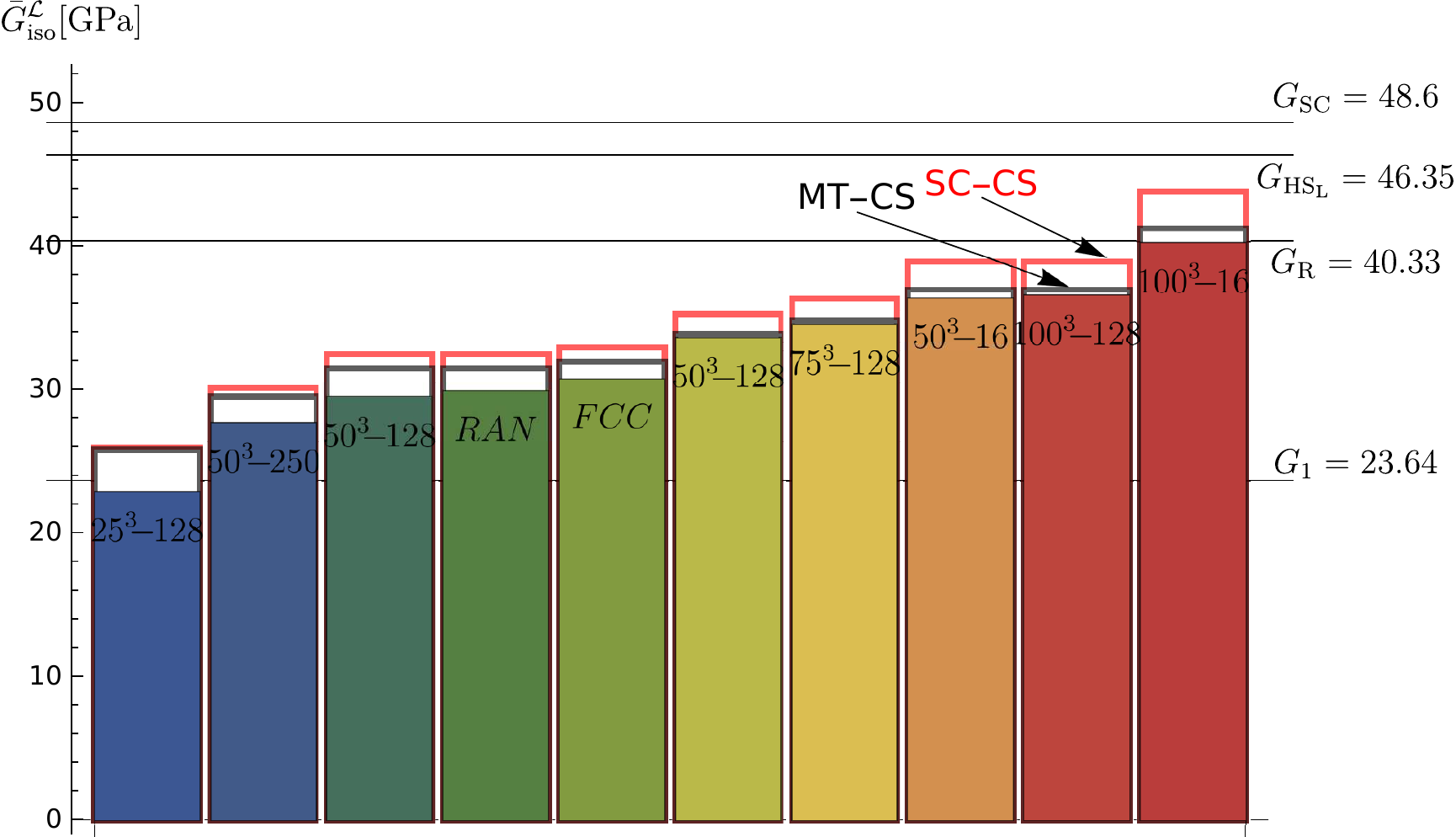}
	\caption{The isotropic shear modulus $\bar{G}^{\mathcal{L}}_{\rm{iso}}$ calculated for the overall stiffness tensor acquired in atomistic simulations for different polycrystalline samples with respect to the selected $\bar{G}$ estimates obtained by continuum mechanics methodology for ideal random polycrystal (horizontal lines: R - Reuss, HS$_{\rm{L}}$ - lower Hashin-Shtrikman bound, SC - self-consistent) and the core-shell model (MT-CS and SC-CS with $\Delta$=5.5\,\AA) estimates for a finite set of orientations (non-filled bars). Samples are ordered according to the increasing value of grain diameter.}
	\label{fig:ShearA}
\end{figure}

\begin{figure}[H]
	\centering
	\includegraphics[width=.8\textwidth]{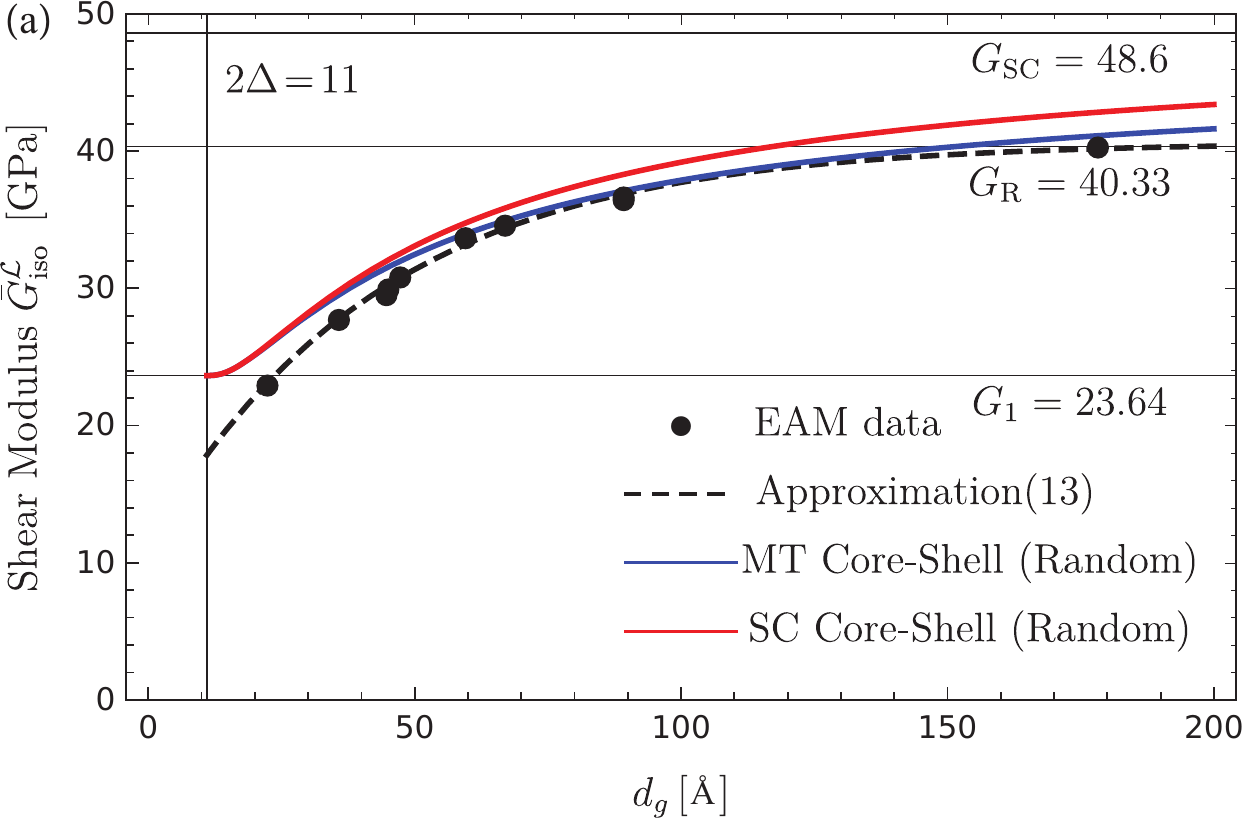}\\
	\includegraphics[width=.8\textwidth]{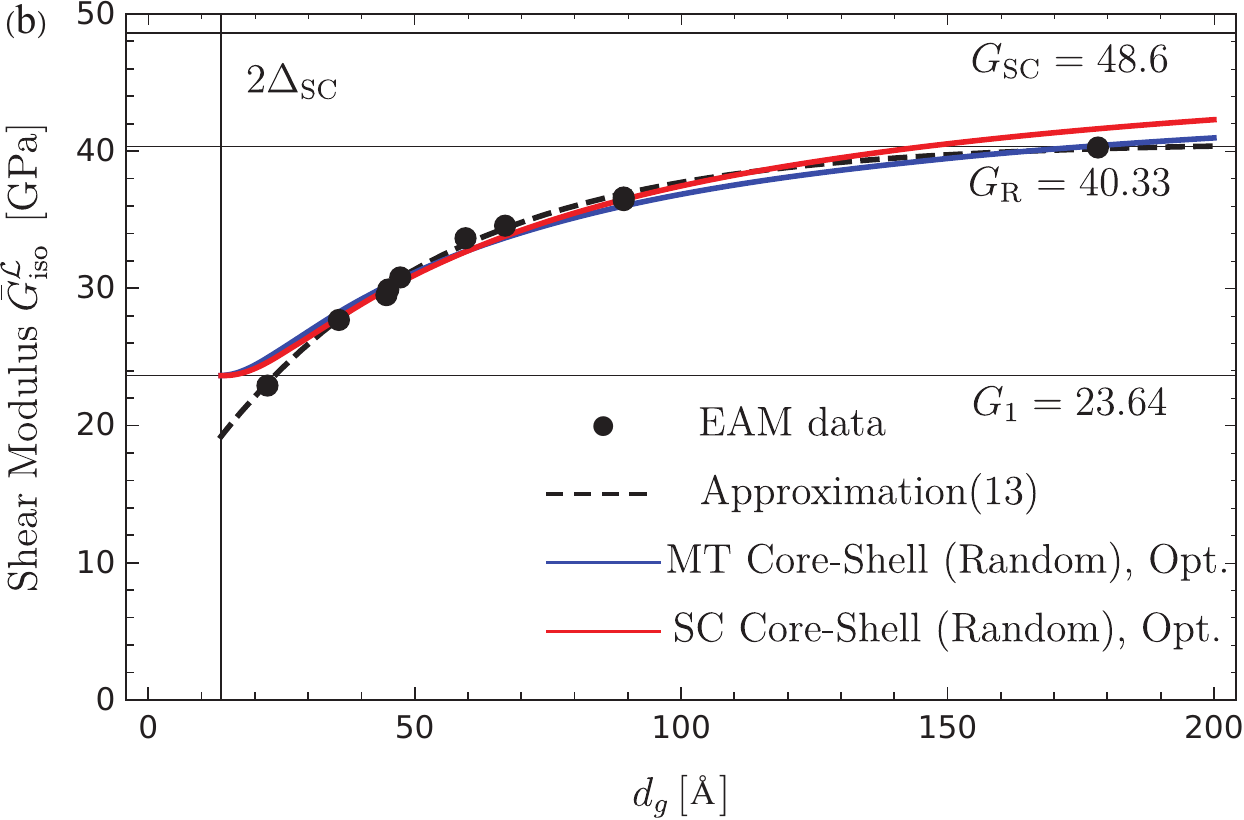}
	\caption{The isotropic shear modulus $\bar{G}^{\mathcal{L}}_{\rm{iso}}$ calculated for the overall stiffness tensor acquired in atomistic simulations as a function of the average grain diameter $d$ for ten polycrystalline samples and by the two variants of the core-shell model. Exponential approximation (\ref{Eq:approx}) of dependency is also shown. (a) $\Delta$=5.5\,{\AA} (b) $\Delta=\Delta_{MT}$=6.387\,{\AA} and $\Delta=\Delta_{SC}$=6.837\,{\AA} for MT and SC variants, respectively.} 
	\label{fig:Approx}
\end{figure}

\begin{figure}[H]
	\centering
	\includegraphics[width=.8\textwidth]{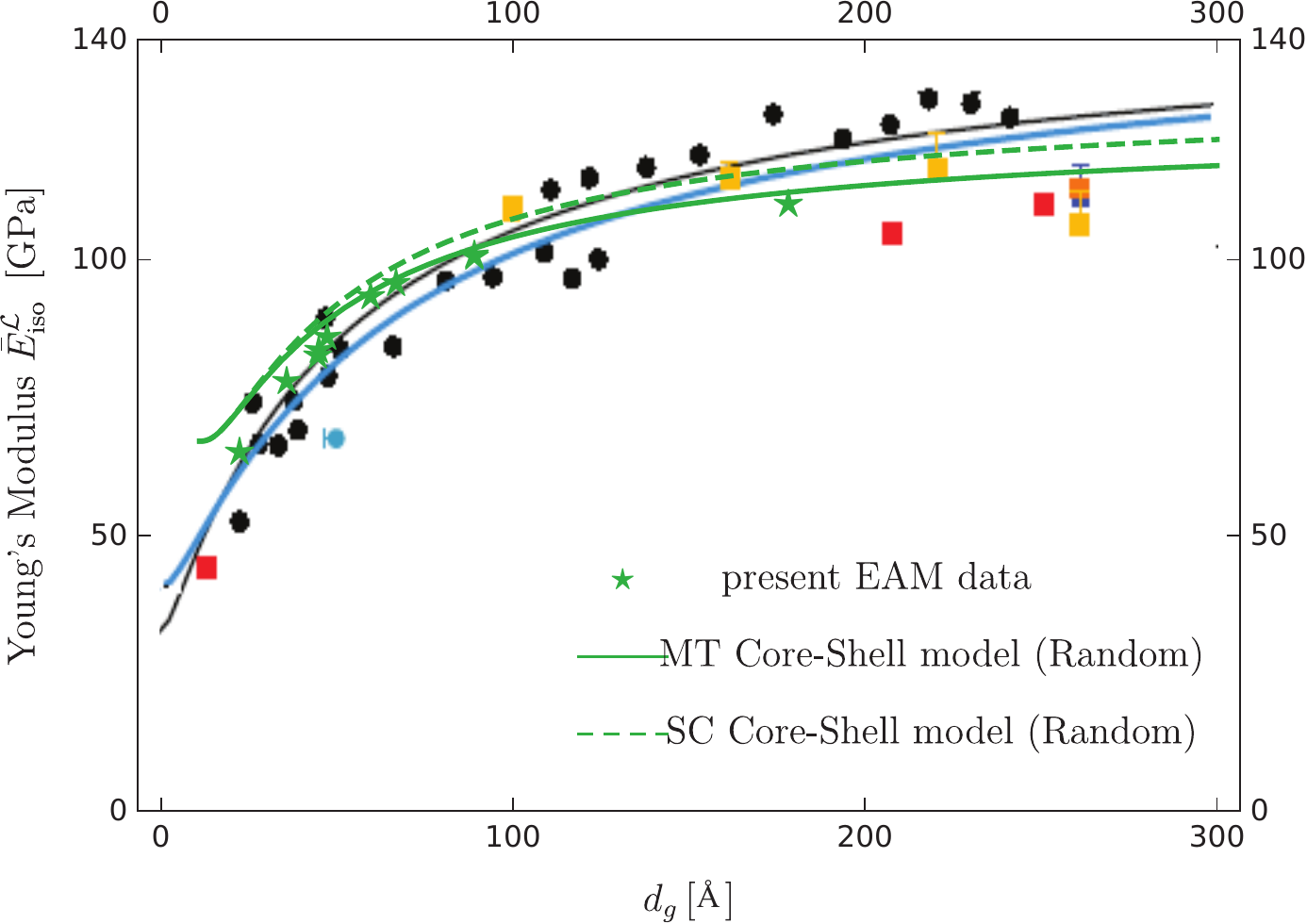}
	\caption{The isotropic Young's modulus $\bar{E}^{\mathcal{L}}_{\rm{iso}}$ calculated for the overall stiffness tensor acquired in atomistic simulations for ten polycrystalline samples (green stars) and by the MT and SC core-shell models with $\Delta$=5.5\,{\AA} (green line) as a function of the average grain diameter $d$. Additionally, results of other experimental and atomistic studies reported in \cite{Gao13} are shown. In particular, black dots represent the results of simulations of \cite{Gao13} (performed at 300\,K), while squares represent the experimental data (for detailed references see \cite{Gao13})}
	\label{fig:ApproxE}
\end{figure}

\begin{figure}[H]
	\centering
	\includegraphics[width=.8\textwidth]{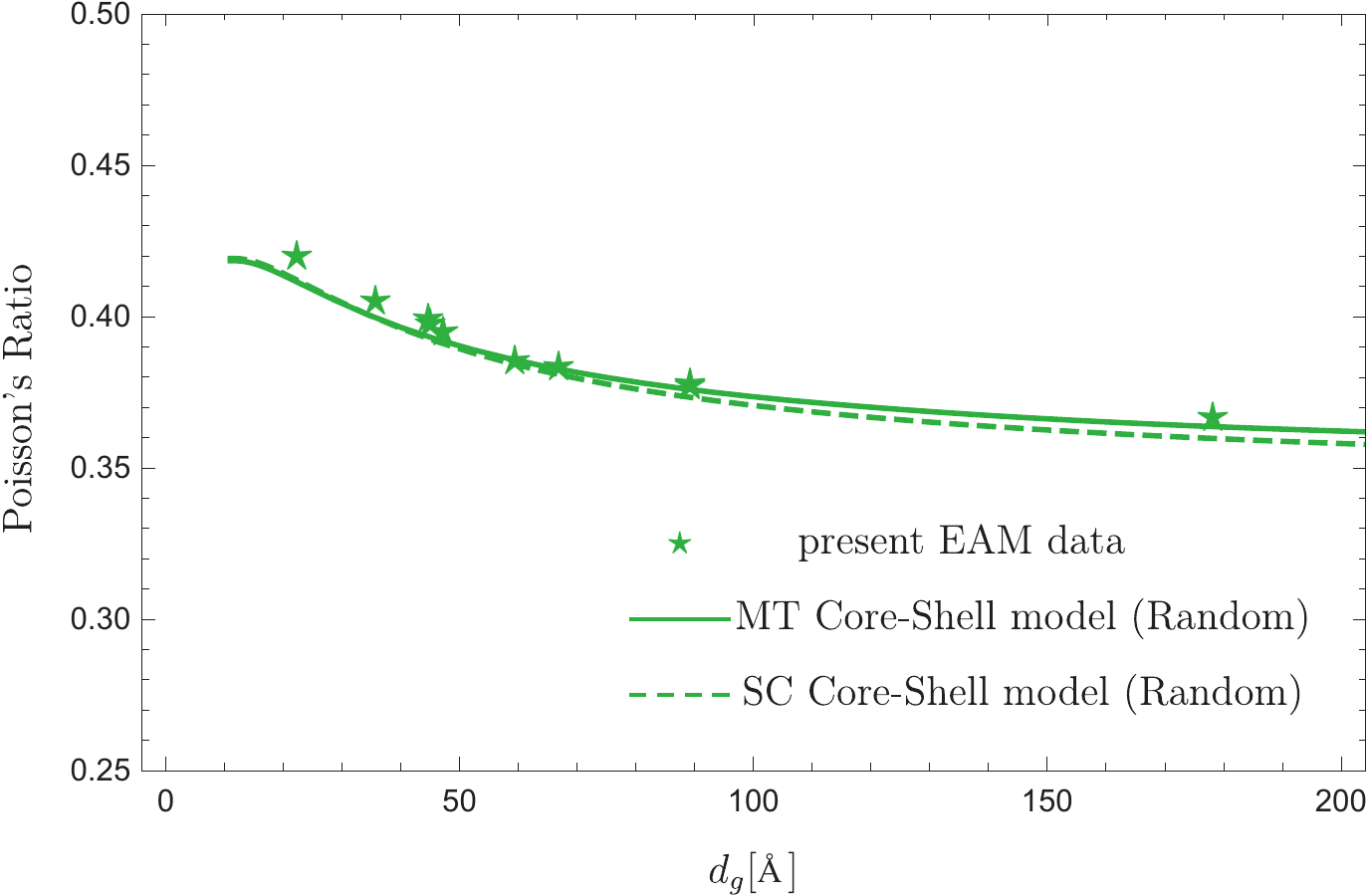}
	\caption{The Poisson's ratio $\bar{\nu}^{\mathcal{L}}_{\rm{iso}}$ calculated for the isotropized overall stiffness tensors acquired in atomistic simulations  for ten polycrystalline samples (green stars) and by the core-shell model (green line) as a function of the average grain diameter $d$.}
	\label{fig:ApproxNu}
\end{figure}
	
As concerns the estimated values of elastic moduli, presented results (see Fig. \ref{fig:BulkA}) indicate that the isotropic bulk modulus $\bar{K}^{\mathcal{L}}_{\rm{iso}}$, established in atomistic calculations for all samples is similar and close to the continuum-mechanics estimate $\bar{K}$=138.36\,GPa. On the contrary, the value of the isotropized shear modulus $\bar{G}^{\mathcal{L}}_{\rm{iso}}$ established in atomistic simulations strongly depends on the grain size. It increases with the average grain diameter used in the simulations, approaching the Reuss continuum-mechanics estimate for the largest grains (see Fig. \ref{fig:ShearA}). The observed dependence can be approximated by the following exponential function:
\begin{equation}\label{Eq:approx}
	\bar{G}^{\mathcal{L}}_{\rm{iso}}(d)=G_0 + (G_{\infty} - G_0) (1 - \exp(-\beta d))
\end{equation}   
where $G_0$ is a shear modulus for 
$d\rightarrow 0$, $G_{\infty}$ - a shear modulus for a coarse-grained polycrystal and $\beta$ governs the rate of change of $\bar{G}^{\mathcal{L}}_{\rm{iso}}$ with the grain size. It is found that $G_0=11.27$\,GPa, $G_{\infty}=40.65$\,GPa and $\beta=0.023$ for the considered copper polycrystal (see Fig. \ref{fig:Approx}). Using the presented results obtained for the bulk and shear moduli the isotropic tensile Young's modulus $\bar{E}^{\mathcal{L}}_{\rm{iso}}$ and Poisson's ratio $\bar{\nu}^{\mathcal{L}}_{\rm{iso}}$ can be derived by well-known formulas: 
\begin{equation}\label{Eq:E-nu}
	\bar{E}^{\mathcal{L}}_{\rm{iso}}=\frac{9\bar{K}^{\mathcal{L}}_{\rm{iso}}\bar{G}^{\mathcal{L}}_{\rm{iso}}}{3\bar{K}^{\mathcal{L}}_{\rm{iso}}+\bar{G}^{\mathcal{L}}_{\rm{iso}}}\,,\quad\bar{\nu}^{\mathcal{L}}_{\rm{iso}}=\frac{3\bar{K}^{\mathcal{L}}_{\rm{iso}}-2\bar{G}^{\mathcal{L}}_{\rm{iso}}}{6\bar{K}^{\mathcal{L}}_{\rm{iso}}+2\bar{G}^{\mathcal{L}}_{\rm{iso}}}\,.
\end{equation}		
The obtained atomistic estimate of Young's modulus shows similar dependence on the grain size as the shear modulus and is in line with the outcomes of other literature studies \cite{Gao13} (see Fig. \ref{fig:ApproxE}). The calculated Poisson's ratio $\bar{\nu}^{\mathcal{L}}_{\rm{iso}}$ decreases with an increasing grain size (Fig. \ref{fig:ApproxNu}). Note that similar predictions concerning Poisson's ratio were obtained by \cite{Kim20123942} with use of the modified continuum theory that includes scale effects and a  non-conforming finite element method.

The tendencies observed in atomistic simulations are well captured, both qualitatively and quantitatively, by two variants of core-shell model formulated in Section \ref{sec:Cont}. Similarly to atomistic results, in the model the bulk modulus does not depend on the grain size and is equal to the local bulk modulus. The results concerning other elastic constants are compared in figures \ref{fig:Approx}-\ref{fig:ApproxNu}.  The volume fraction of the transient zone $f_0$ was calculated using the \textit{cutoff} radius of atomistic potential (Eq. (\ref{def:fsa})). Although the specific values of $\bar{G}$ and $\bar{E}$ are slightly over-predicted for very small grains, the quality of obtained estimates is satisfactory in the whole range of grain sizes. It is worth to recall and underline that no fitting of model parameters was performed to achieve this remarkable agreement, therefore such selection of $\Delta$ can be recommended for the preliminary assessment of effective elastic properties of nanocrystalline copper. Even better agreement is obtained if the value $\Delta$ in Eq. (\ref{def:fsa}) is established to fit best the atomistic results by Eq. (\ref{eq:core-shell}). These optimized values are found to be 6.387\,{\AA} for MT and 6.837\,{\AA} for the SC variant, respectively (see Fig. \ref{fig:Approx}b).
	
\section{Conclusions}
\label{sec:Con}

The paper discusses the applicability of continuum-mechanics mean-field one-phase and two-phase models to estimating effective elastic properties of bulk nanocrystalline FCC copper. The methods used to verify such micromechanical mean-field estimates in the case of coarse-grained materials, such as full-field simulations on representative aggregates using the FEM or FFT framework, when they are combined with the classical linearly elastic constitutive law, are not applicable in the present context. Therefore, in this paper a set of atomistic simulations have been performed on the generated grain aggregates with randomly selected orientations.  At variance with the available literature, all 21 components of the elasticity tensor $\bar{\mathbb{C}}$  are acquired by performing six independent numerical tests for ten samples of the polycrystalline material. The obtained results are then analysed with respect to the anisotropy degree of the overall stiffness tensor $\bar{\mathbb{C}}$, resulting from the limited number of grain orientations and their spatial distribution. The closest isotropic approximation of this tensor is found using the Log-Euclidean norm \cite{Moakher06}. The dependence of the obtained overall bulk and shear moduli on the average grain diameter is analysed. It is found that, while the shear modulus decreases with the grain size, the bulk modulus shows negligible dependence on the grain diameter and is close to the bulk modulus of a single crystal. When another pair of parameters: Young's modulus and Poisson's ratio are calculated, it is found that with a decreasing grain size the former decreases (in agreement with other numerical and experimental studies), while the latter increases. 

Two variants of an anisotropic two-phase model of a nano-grained polycrystal, built in the spirit of \cite{Jiang04,Capolungo07} and called MT and SC core-shell models, have been formulated. In the model the thickness of the shell is specified by the \emph{cutoff radius} of a corresponding atomistic potential, while the grain shell has the stiffness tensor corresponding to the lower zero-order bound of $\bar{\mathbb{C}}$. Under such assumptions, in the case of grain cores with cubic elastic symmetry, the effective stiffness tensor of a bulk polycrystal is specified by an explicit formula. It has been shown that the obtained estimates are in satisfactory qualitative and quantitative agreement with the results of atomistic simulations performed for nano-crystalline copper. In particular, in accordance with the atomistic simulations, the predicted bulk modulus does not depend on the grain size, while the shear modulus decreases with it.

In future studies it should be checked whether the observed difference in the dependence of the bulk and shear moduli on the grain size is valid for other metals with cubic lattice symmetry, especially when they differ in their Zener anisotropy factor. Such studies have been already initiated. The proposed mean-field two-phase model can be extended to estimate a non-linear response of a nano-grained polycrystal (compare \cite{Jiang04,Capolungo07}) and specifically the yield strength. The extension will also require verification based on atomistic simulations.

 
\appendix

\section{Classical estimates of effective properties of one-phase polycrystal}
\label{Ap:A}

The following classical estimates of effective properties of one-phase polycrystal are considered in this paper as reference values:
\begin{itemize}
	\item The Voigt and Reuss bounds equivalent to the uniform strain and uniform stress assumption \cite{Hill52} and independent of the grain shape:
	\begin{equation}\label{lowerL}
	\bar{\mathbb{C}}_{\rm{V}}=\left<\mathbb{C}(\phi^c)\right>,\quad
	\bar{\mathbb{C}}_{\rm{R}}=\left<\mathbb{S}(\phi^c)\right>^{-1}\,.
	\end{equation}
	
	\item The bounds resulting from the Hashin-Shtrikman variational principle \cite{Hashin62a,Hashin62b}:
	\begin{equation}\label{HSbound}
	\bar{\mathbb{C}}_{\rm{HS}}=\left<(\mathbb{C}(\phi^c)+\mathbb{C}_*(\mathbb{C}_0))^{-1}\right>^{-1}-\mathbb{C}_*(\mathbb{C}_0)\,,
	\end{equation}
	where $\mathbb{C}_*(\mathbb{C}_0)$ is the Hill tensor \cite{Hill65} depending on
	the stiffness $\mathbb{C}_0$ of comparison
	material and the shape of the equivalent inclusion (here assumed as spherical). In order to obtain the upper $\bar{\mathbb{C}}_{\rm{HS}}^{\rm{U}}$ and the lower $\bar{\mathbb{C}}_{\rm{HS}}^{\rm{L}}$ bounds of the stiffness tensor the $\mathbb{C}_0$ is selected in a way to ensure the positive or negative definiteness of the tensor
	$\mathbb{C}_0-\mathbb{C}(\phi^c)$ for any $\phi^c$, respectively \cite{Walpole81,Berryman05,Kowalczyk12}. In this paper the bounds are specified for isotropic comparison material. The bulk and shear moduli of optimal $\mathbb{C}_0$ are called 0-rth order bounds for random polycrystal \cite{Nadeau01}.
	
	\item The self-consistent estimate obtained by approximating the strain in each grain as equal to the strain in a spherical inclusion with $\mathbb{C}(\phi_c)$-stiffness embedded in an infinite medium of effective properties to be found \cite{Kroner58,Hill65}, namely
	\begin{equation}\label{sc1}
	\bar{\mathbb{C}}_{\rm{SC}}=\left<(\mathbb{C}(\phi^c)+\mathbb{C}_*(\bar{\mathbb{C}}_{\rm{SC}}))^{-1}\right>^{-1}-\mathbb{C}_*(\bar{\mathbb{C}}_{\rm{SC}})\,.
	\end{equation}	
	The Eshelby result \cite{Eshelby57} is used to obtain the above formula. The formula is implicit.
\end{itemize}
Note that in the case of one-phase polycrystal averaging over the representative volume is equivalent to the averaging over the grain orientations according to Eq. (\ref{Eq:orient}).  For the crystal of cubic symmetry and random orientation distribution the respective expressions for the shear modulus $\bar{G}$ obtained using the above models are collected in Table \ref{tab:LinearEstimates}.

\begin{table}[!htp]
	\caption{Continuum-mechanics estimates of the overall shear modulus $\bar{G}$\,[GPa] for an elastic random cubic polycrystal. The respective quantitative predictions for a copper polycrystal are calculated for a monocrystal elastic stiffness (see Table \ref{tab:Cij}) for which: $K=138.36$\,GPa, $G_1=23.64$\,GPa, $G_2=76.19$\,GPa (see Eq. (\ref{Eq:moduli})).
	}
	\label{tab:LinearEstimates}\vspace{.05in}
	\centering
	\small{
		\begin{tabular}{|c|c|c|}
			\hline
			Estimate    & Expression & $\bar{G}$ for copper   \\
			
			& $\bar{G}$ & random polycrystal  \\
			\hline
			Voigt (V)    & $\frac{1}{5}\left(2G_1+3G_2\right)$ &  55.17 \\
			\hline
			&&\\
			Reuss (R) & {\Large{$\frac{5G_1G_2}{3G_1+2G_2}$}}    &   40.33   \\
			&&\\
			\hline
			&&\\
			Hashin-Shtrikman (SH) & {\Large{$\frac{G_{\ast}^{U}(2G_1+3G_2)+5G_1G_2}{5G_{\ast}^{U}+3G_1+2G_2}$}} & 49.90\\
			UPPER  &  {\small{$G_{\ast}^{U}=G_0^U\frac{9K+8G_0^U}{6(K+2G_0^U)}$}}    &   \\
			& $G_0^U=\rm{max}(G_1,G_2)$ &\\
			\hline
			&&\\
			Hashin-Shtrikman (HS) &  {\Large{ $\frac{G_{\ast}^{L}(2G_1+3G_2)+5G_1G_2}{5G_{\ast}^{L}+3G_1+2G_2}$}}  &  46.35\\
			LOWER  &   {\small{$G_{\ast}^{L}=G_0^L\frac{9K+8G_0^L}{6(K+2G_0^L)}$}}     &   \\
			& $G_0^L=\rm{min}(G_1,G_2)$ &\\
			\hline
			&single positive root of equation:&\\
			
			Self-consistent (SC)    & $8(\bar{G}^{\rm{SC}})^3+(9K+4G_1)(\bar{G}^{\rm{SC}})^2+$& 48.60\\
			&$-3G_2(4G_1+K)\bar{G}^{\rm{SC}}-6G_1G_2K=0$ &\\
			&&\\
			\hline
		\end{tabular}\\[.051in]
	}
\end{table}

\newpage

\section{Notation}
 
The following array presents the relation between the components of elasticity tensor written in the Voigt notation and as components of the fourth order tensor
 
\begin{eqnarray}
\centering
\left[C_{KL}\right]=\left[
\begin{array}{cccccc}
{C_{1111}} & {C_{1122}} & {C_{1133}} & {C_{1123}} & {C_{1131}} & {C_{1112}} \\
{C_{1122}} & {C_{2222}} & {C_{2233}} & {C_{2223}} & {C_{2231}} & {C_{2212}} \\
{C_{1133}} & {C_{2233}} & {C_{3333}} & {C_{3323}} & {C_{3331}} & {C_{3312}} \\
{C_{1123}} & {C_{2223}} & {C_{3323}} & {C_{2323}} & {C_{2331}} & {C_{2312}} \\
{C_{1131}} & {C_{2231}} & {C_{3331}} & {C_{2331}} & {C_{3131}} & {C_{3112}} \\
{C_{1112}} & {C_{2212}} & {C_{3312}} & {C_{2312}} & {C_{3112}} & {C_{1212}} \\
\end{array}
\right]\,.
\label{eqn:CuCij}
\end{eqnarray} 
For the tensor of cubic symmetry we have three independent components, namely
\begin{equation}
C_{11}=C_{1111}\,,\quad C_{12}=C_{1122}\,,\quad C_{44}=C_{2323}\,.
\end{equation}
 
\section*{References}

\bibliography{References}

\end{document}